\documentclass{emulateapj}

\usepackage{amssymb}
\usepackage{amsmath}
\usepackage{array}
\newcolumntype{C}[1]{>{\centering\let\newline\\\arraybackslash\hspace{0pt}}m{#1}}
\usepackage{graphicx}
\usepackage{natbib}
\usepackage{txfonts}
\usepackage{microtype}
\usepackage{subfigure}
\usepackage[colorlinks=true,citecolor=blue,linkcolor=blue]{hyperref}

\bibpunct{(}{)}{;}{a}{}{,}

\def\roma{1}
\def\icra{2}
\def\rio{3}
\def\exeter{4}
\def\bcna{5}
\def\bcnb{6}

\shorttitle{Spin evolution of magnetized super-Chandrasekhar white dwarfs}

\shortauthors{Becerra et al.}

\begin{document}

\title{The    Spin    Evolution     of    Fast-Rotating,    Magnetized
  Super-Chandrasekhar  White Dwarfs  in the  Aftermath of  White Dwarf
  Mergers}

\author{L.~Becerra\altaffilmark{\roma,\icra},
        J.~A.~Rueda\altaffilmark{\roma,\icra,\rio}
        P.~Lor\'en-Aguilar\altaffilmark{\exeter}, and
		E.~Garc\'\i a--Berro\altaffilmark{\bcna,\bcnb}}
	
\altaffiltext{\roma}{Dipartimento di Fisica, 
                     Sapienza Universit\`a di Roma, 
                     P.le Aldo Moro 5, 
                     I--00185 Rome, 
                     Italy}
                     
\altaffiltext{\icra}{ICRANet, 
                     P.zza della Repubblica 10, 
                     I--65122 Pescara, 
                     Italy} 

\altaffiltext{\rio}{ICRANet-Rio, 
                    Centro Brasileiro de Pesquisas F\'isicas, 
                    Rua Dr. Xavier Sigaud 150, 
                    22290--180 Rio de Janeiro, 
                    Brazil}                     
                     
\altaffiltext{\exeter}{School of Physics, 
                       University of Exeter, 
                       Stocker Road, 
                       Exeter EX4 4QL, 
                       UK}

\altaffiltext{\bcna}{Departament de F\'{i}sica, 
                     Universitat Polit\`{e}cnica de Catalunya, 
                     c/Esteve Terrades, 5, 
                     08860 Castelldefels, 
                     Spain}

\altaffiltext{\bcnb}{Institut d'Estudis Espacials de Catalunya, 
                     Ed. Nexus-201, 
                     c/Gran Capit\`a 2-4, 
                     08034 Barcelona, 
                     Spain}

\begin{abstract}
The evolution  of the  remnant of  the merger of  two white  dwarfs is
still an  open problem.  Furthermore,  few studies have  addressed the
case in which the remnant is a magnetic white dwarf with a mass larger
than the  Chandrasekhar limiting mass.  Angular  momentum losses might
bring the  remnant of the  merger to the physical  conditions suitable
for developing  a thermonuclear explosion. Alternatively,  the remnant
may be  prone to gravitational or  rotational instabilities, depending
on  the  initial conditions  reached  after  the coalescence.   Dipole
magnetic braking is  one of the mechanisms that can  drive such losses
of angular  momentum.  However,  the timescale  on which  these losses
occur depend on several parameters,  like the strength of the magnetic
field.  In  addition, the  coalescence leaves a  surrounding Keplerian
disk that  can be accreted by  the newly formed white  dwarf.  Here we
compute the post-merger evolution  of a super-Chandrasekhar magnetized
white dwarf taking  into account all the  relevant physical processes.
These include magnetic torques acting  on the star, accretion from the
Keplerian disk, the threading of the magnetic field lines through the
disk, as  well as the thermal  evolution of the white  dwarf core.  We
find that  the central  remnant can reach  the conditions  suitable to
develop a thermonuclear explosion  before other instabilities (such as
the  inverse  beta-decay  instability   or  the  secular  axisymmetric
instability) are  reached, which  would instead lead  to gravitational
collapse of the magnetized remnant.
\end{abstract}


\maketitle 

%%%%%%%%%%%%%%%%%%%%%%%%%%%%%%%%%%%%%%%%%%%%%%%%%
%%%%%%%%%%%%%%%%%%%%%%%%%%%%%%%%%%%%%%%%%%%%%%%%%
\section{Introduction}
%%%%%%%%%%%%%%%%%%%%%%%%%%%%%%%%%%%%%%%%%%%%%%%%%
%%%%%%%%%%%%%%%%%%%%%%%%%%%%%%%%%%%%%%%%%%%%%%%%%

Type Ia  supernovae (SNe~Ia) are  one of the most  energetic explosive
events in the cosmos.  Since  there is a well established relationship
between their  intrinsic brightnesses  and the  shapes of  their light
curves \citep{Phillips}  and, moreover,  because they are  so luminous
that they can be detected at  very large distances, SNe~Ia can be used
as as standardizable cosmological candles.   This has opened a new era
in cosmology,  and has enabled us  to measure the acceleration  of the
universe  \citep{Riess,Perlmutter}.   Additionally,   SNe~Ia  play  an
important role in  modern cosmology, since they allow us  to probe the
fundamental  physical  theories underlying  dark  energy  -- see,  for
instance,  \citet{Weinberg} for  a  recent review  on this  interesting
topic.

Nevertheless, despite the  importance of SNe~Ia, we still  do not know
the nature of their progenitor systems.  Actually, this has remained a
long-standing  mystery  for various  decades.   We  do know  that  the
outburst is driven by the explosion  of a carbon-oxygen white dwarf in
a  binary system,  but  we  do not  know  the  precise mechanism  that
destabilizes  the  white  dwarf.   Several hypothesis  have  been  put
forward over the years, and most likely SNe~Ia may have a diversity of
progenitors. In  the following we superficially  describe the possible
evolutionary  channels   leading  to   a  SN~Ia.   In   the  so-called
single-degenerate  channel  (SD) a  white  dwarf  in a  binary  system
accretes matter  from a non-degenerate  companion, and explodes  as it
approaches  the  Chandrasekhar limiting  mass  --  see, for  instance,
\citet{Han2004}  for a  recent  discussion.   In the  double-degenerate
scenario (DD) two  white dwarfs members of a close  binary system lose
angular  momentum and  energy through  the radiation  of gravitational
waves,  and  a  merger   occurs  \citep{Webbink1984,  IT84}.   Another
possible scenario  is the core-degenerate scenario  (CD).  Within this
formation  channel a  hot core  is  formed at  the end  of the  common
envelope  phase  of the  binary  system  \citep{Livio,Soker}, and  the
merger of  the core of  the asymptotic giant  branch (AGB) star  and a
secondary    white    dwarf    powers    the    explosion    --    see
\cite{aznarsiguanetal15}  for  a  simulation of  the  merger  process.
Another  recently  proposed pathway  is  the  white dwarf-white  dwarf
collisional  scenario,  in  which  two white  dwarfs  collide  --  see
\citet{Aznarsiguan2013} for a recent summary of the relevant literature
on the subject.  Despite some attractive features of this scenario, it
has been shown that it can only account  for at most a few per cent of
all SNe~Ia  \citep{Ilkov}.  Each of  these formation channels  has its
own  advantages   and  drawbacks,   and  in   some  cases   these  are
severe. Currently,  one of the  most favored  ones is the  DD channel,
because it  provides adequate  answers to two  important observational
facts,  namely   the  absence  of   hydrogen  in  the   nebular  phase
\citep{Leonard},  and  the  delay  time  distribution  \citep{Totani}.
Consequently, we focus on it.

Smoothed-Particle-Hydrodynamics (SPH) simulations of  the coalescence of binary white dwarfs show that a prompt explosion is not always the result of  the interaction  during the dynamical  phase of the merger \citep{2007MNRAS.380..933Y,2009A&A...500.1193L}.
Only those binary  systems in which both
the secondary  and primary stars  are massive  enough lead to  a SN~Ia
outburst. Actually,  the parameter space  for violent mergers  is very
narrow \citep{pakmor,Sato15},  and only  when two  carbon-oxygen white
dwarfs of masses larger than $\sim 0.8\, M_{\sun}$ merge the result of
the dynamical  phase is a  prompt explosion.  Within  this theoretical
framework, the dynamical  phase of the merger is followed  by a second
phase in which  the material of the debris region  can be accreted and
possibly lead to a SN~Ia explosion.

The existing  simulations of  the coalescence  of binary  white dwarfs
\citep{1990ApJ...348..647B,  2009A&A...500.1193L, 2012ApJ...746...62R,
2013ApJ...767..164Z, 2014MNRAS.438...14D}  show that the  final result
of  the coalescence  consists of  a central  white dwarf  made of  the
undisrupted primary  star and a  hot, convective corona made  of about
half of the mass of the  disrupted secondary.  This central remnant is
surrounded by a heavy Keplerian disk, made  of the rest of the mass of
the disrupted secondary, since little  mass is ejected from the system
during  the  merger  episode.   The rapidly-rotating,  hot  corona  is
convective and an efficient $\alpha\omega$ dynamo can produce magnetic
fields  of  up  to $B\approx  10^{10}$~G  \citep{2012ApJ...749...25G}.
Recent  two-dimensional  magneto-hydrodynamic  simulations  of
  post-merger systems confirm  the growth of the  white dwarf magnetic
  field after  the merger \citep{Ji,2015ApJ...806L...1Z}.  Thus, the
role of magnetic fields in the aftermath of the dynamical merger needs
to be explored. This is precisely the aim of the present work, as very
few   studies  \citep{Ji,Beloborodov14,kulebi}   of  the   post-merger
evolution including  magnetism has been  done so far.   On the
  other hand,  there are few works  that explore the evolution  of the
  post-merger  systems  but without  considering  the  effects of  the
  central      remnant      magnetic-field     --      see,      e.g.,
  \cite{2007MNRAS.380..933Y}, \cite{vk10}, \cite{2012ApJ...748...35S},
  and \cite{2012MNRAS.427..190S}.

Our     paper     is     organized     as     follows.      In
  Section~\ref{sec:post-merger} we explain the model of the evolution of
  the post-merger system.  In particular, we describe the torques that
  act on  the central remnant (Section~\ref{sec:torque}),  the structure
  of  the rotating  central white  dwarf (Sec.~\ref{sec:WD}),  and its
  thermal evolution  (Section~\ref{sec:TWD}).  In Section~\ref{sec:mdotWD}
  we present the prescriptions adopted  to model the accretion rate on
  the central star.  In  Section~\ref{sec:initial} we present our choice
  of initial conditions. Later, in Section~\ref{sec:results}, we discuss
  extensively the results of our  simulations for two prescriptions to
  compute the accretion  rate.  Specifically, in Section~\ref{sec:tcool}
  we  discuss  the  case  in  which the  accretion  rate  is  obtained
  employing the  cooling timescale, while in  Section~\ref{sec:tvisc} we
  present the results  obtained when it is computed  using the viscous
  timescale. Finally, in Sec.~\ref{sec:comparison} we compare our results with previous works and in Sec.~\ref{sec:concl} we summarize our major findings, we elaborate on their  significance and  we present  our
  concluding remarks.

%%%%%%%%%%%%%%%%%%%%%%%%%%%%%%%%%%%%%%%%%%%%%%%%%
%%%%%%%%%%%%%%%%%%%%%%%%%%%%%%%%%%%%%%%%%%%%%%%%%
\section{Numerical setup for the post-merger evolution}
\label{sec:post-merger}
%%%%%%%%%%%%%%%%%%%%%%%%%%%%%%%%%%%%%%%%%%%%%%%%%
%%%%%%%%%%%%%%%%%%%%%%%%%%%%%%%%%%%%%%%%%%%%%%%%%

Numerical simulations of  binary white dwarf mergers  indicate that in those  cases in  which  a violent  merger does  not  occur the  merged
configuration has three distinct regions \citep{1990ApJ...348..647B, 2004A&A...413..257G, 2009A&A...500.1193L, 2012ApJ...746...62R,
2013ApJ...767..164Z, 2014MNRAS.438...14D}.
First, there is the central white dwarf that rotates as a rigid solid.  On top of it  a hot,  differentially-rotating, convective  corona can  be found. This corona  is made of  matter accreted from the  disrupted secondary star.   The mass  of this  region is  about half  of the  mass of  the secondary white  dwarf.  Finally, surrounding these  two regions there
is a rapidly rotating Keplerian disk, which is made of the rest of the
material of the disrupted secondary, since only a small amount of mass
is ejected from  the system during the coalescence.   The evolution of
the post-merger configuration  depends, naturally, on the  mass of the
central remnant, which is made of  the primary white dwarf and the hot
corona, and on the properties of the surrounding disk.

Before  entering  into  details  it   is  important  to  discuss  some
timescales, which are relevant to adopt a reasonable approximation for
the evolution of the system.   The post-merger configuration formed in
the coalescence has a clear  hierarchy of timescales.  First comes the
dynamical timescale $t_{\rm dyn}\sim  \Omega^{-1}$, being $\Omega$ the
rotation velocity. This  timescale is typically of the order  of a few
seconds.  Next in order of magnitude is the viscous timescale, $t_{\rm
visc}$, of the  Keplerian disk.  This timescale  governs the transport
of disk  mass inwards and  of angular momentum outwards.   The viscous
phase of the evolution is  normally followed using the Shakura-Sunyaev
$\alpha$ prescription   \citep{SS73}.    Adopting   typical   values
\citet{vk10}  found  that $t_{\rm  visc}$  ranges  between $10^3$  and
$10^4$~s.  However, for the merger  remnants studied here we find that
the viscous  timescales could be significantly  shorter, typically 1~s
-- see  Sec.~\ref{sec:disk}. These viscous timescales are a little longer, but of the same order, of the dynamical timescale of the post-merger, super-Chandrasekhar white dwarf studied here ($t_{\rm dyn}\sim 0.2-0.3$~s, see Sec.~\ref{sec:WD}). Finally,  in this  set of  characteristic times  is the
thermal  timescale  of the  merger  product,  $t_{\rm th}$,  which  is
typically  much longer  and  will  be discussed  in  detail for  every
simulation.

In the following we first discuss the interactions between the
  remnant and  the disk in order  to identify the torque  that acts on
  the  central  remnant  and  model   the  evolution  of  its  angular
  momentum. Then we present how we model the structure of the rotating
  white dwarf central  remnant, as well as its  thermal evolution. The
  fate of the  white dwarf depends crucially on the  accretion rate in
  the post-merger phase. Thus, we consider two very different physical
  scenarios  to   determine  the  accretion  rate   onto  the  central
  remnant. In  the first of them  the accretion rate is  determined by
  the viscous time-scale, whereas in the second scenario the accretion
  rate is governed  by thermal time-scale. These will  be discussed in
  Section~\ref{sec:mdotWD}.

%%%%%%%%%%%%%%%%%%%%%%%%%%%%%%%%%%%%%%%%%%%%%%%%%
\subsection{Torques on the central remnant}
\label{sec:torque}
%%%%%%%%%%%%%%%%%%%%%%%%%%%%%%%%%%%%%%%%%%%%%%%%%

The  most  commonly  employed  model  of disk  evolution  is  that  of
\cite{1979ApJ...232..259G}.    This  model   was  later   improved  by
\cite{1987A&A...183..257W},  and  \cite{1995ApJ...449L.153W}.   Within
this model the magnetic field of  the remnant penetrates the disk in a
broad   transition  zone   as   a  result   of  the   Kelvin-Helmholtz
instabilities,  turbulent diffusion  and magnetic  field reconnection.
Furthermore, the material of the disk  corotates with the star only in
a narrow  region, and  reaches its surface  channeled by  the magnetic
field  lines.   We assume  that  the  magnetic  field of  the  remnant
penetrates  the   disk  up   to  approximately  the   Alfv\'en  radius
\citep{1973ApJ...179..585D}:
\begin{equation}
  R_{\rm mag}=\left(\frac{ \mu_{\rm WD}^2}{ \dot{M}_{\rm disk}\sqrt{\,2GM_{\rm WD}}}
  \right)^{2/7},
\label{eq:Rmag}
\end{equation}
where $\mu_{\rm WD}=B_{\rm WD}R_{\rm WD}^3$  is the magnetic moment of
the star, $B_{\rm WD}$ its magnetic field, and $\dot{M}_{\rm disk}$ is
the mass flow through the inner radius of the disk, $R_0$.

The angular momentum per unit  mass entering into the magnetosphere of
the white  dwarf through  the inner radius  is $l_{0}=R_0^2\Omega_{\rm
K}^{0}$, where
\begin{equation}
\Omega_{\rm K}^{0}=\left(\frac{GM_{\rm WD}}{R_0^3} \right)^{1/2}
\end{equation}
is the Keplerian angular velocity at $R_0$. This material is channeled
by  the magnetic  field onto  the surface  of the  remnant. Thus,  the
resulting spin-up torque  on the star due accretion of  disk matter is
given by:
\begin{equation}
T_{\rm acc}=\xi_{\rm acc}\dot{M}_{\rm WD}R_0^2\Omega_{\rm K}(R_0),
\label{eq:Tacc}
\end{equation}
where  $\dot{M}_{\rm WD}=\varepsilon\dot{M}_{\rm  disk}$  is the  mass
accretion  rate  on  the   white  dwarf,  $\varepsilon$  measures  how
efficient  accretion  is, and  $\xi_{\rm  acc}$  is a  parameter  that
accounts  for the  deceleration of  the accreted  matter in  the inner
region of the  disk. The values adopted for these  two parameters will
be discussed below.
If the star  rotates faster than the matter at  the inner disk radius,
the centrifugal barrier blocks this  material.  Hence, it cannot reach
the surface  of the newly born  white dwarf. This matter  is therefore
ejected from the system. This happens  when the inner edge of the disk
moves beyond  the corotation  radius, that is  beyond the  distance at
which the disk  rotates with the same angular velocity  of the central
object:
\begin{equation}
  R_{\rm co}=\left(\frac{GM_{\rm WD}}{\Omega_{\rm WD}^2}\right)^{1/3}.
\end{equation}
When  $R_0>R_{\rm  co}$,  the  system  enters  into  the  dubbed  {\sl
propeller regime} \citep{1975A&A....39..185I}.  During this phase, the
material reaching  the magnetosphere  is ejected with  higher specific
angular momentum that  the one it had previously, thus  resulting in a
spin-down  torque  acting  on  the  star.  This  torque  is  given  by
\citep{1999ApJ...520..276M}:
\begin{equation}\label{eq:Tprop}
T_{\rm prop}=\sqrt{GM_{\rm WD}R_{0}} \dot{M}_{\rm disk}\left(1-\omega_{\rm f}\right),
\end{equation}
In this expression we have introduced the so-called {\sl fastness parameter}:
\begin{equation}\label{eq:fastness}
  \omega_{\rm f}=\Omega_{\rm WD}/\Omega_{\rm K}(R_0)=
  \left( R_0/R_{\rm co} \right)^{3/2}.
\end{equation}

The rotating magnetic field of the star originates an induced electric
field that  results in a  wind. The  wind fills the  magnetosphere and
corotates   with   the   star  \citep{1969ApJ...157..869G}.   At   the
light-cylinder,   $R_{\rm  lc}=c/\Omega_{\rm   WD}$,  the   corotation
velocity  reaches the  speed of  light, delimiting  the region  within
which  corotation   of  the  magnetosphere  is   enforced.  At  larger
distances, the field lines become  open.  For the spin-down torque due
to  electromagnetic  energy  losses  in  a  force-free  magnetosphere,
$T_{\rm dip}$, we adopt the result of \cite{2006ApJ...648L..51S}:
\begin{equation}\label{eq:Tdipole}
T_{\rm dipole}=k_1\frac{\mu_{\rm WD}^2\Omega^3_{\rm WD}}{c^3}(1+k_2\sin^2{\theta}),
\end{equation}
with $k_1=1\pm  0.05$ and  $k_2=1\pm 0.1$, and  $\theta$ is  the angle
between the magnetic moment and the rotation axis of the star.

Finally,  an  additional  torque,  $T_{\rm  mag}$,  results  from  the
interaction of  the disk with the  magnetic field of the  white dwarf.
According to \cite{1979ApJ...232..259G}, matter  of the disk moving in
the magnetic field of the  white dwarf generates currents that confine
the stellar field  inside the disk.  The magnetic  field threading the
disk  is $\vec{B}_{\rm  disk}=\eta  \vec{B}_{\rm  WD}^{\rm p}$,  where
$\vec{B}_{\rm  WD}$  is  the  magnetic   field  of  the  white  dwarf,
$\vec{B}_{\rm WD}^{\rm p}$ is its projection on the plane of the disk,
and $\eta\leq 1$  is the screening coefficient which  accounts for the
effect of  currents in the  partially diamagnetic disk induced  by the
stellar field.  In particular, the poloidal field induces an azimuthal
current due to the radial motion of the plasma, that partially screens
the stellar  magnetic field.   Also, the  relative motion  between the
disk and the star magnetosphere  generates a toroidal field, $B_\phi$.
In the simulations, we have considered  that the growth of $B_\phi$ is
limited by  diffusive decay  due to turbulent  mixing within  the disk
\citep{1995ApJ...449L.153W}.  Following the  analytical formulation of
\cite{1997ApJ...475L.135W}, the magnetic torque  acting on the central
star due to its interaction with the disk is given by:
\begin{eqnarray}
\label{eq:Tdint}
T_{\rm mag}&=&\frac{\Gamma \eta^2 \mu_{\rm WD}^2}{R_0^3}\nonumber\\
   &~&\left[\frac{2h}{R_0}(1-\omega_f)\sin^2\theta
  +\frac{\cos^2{\theta}}{3}(1-2\omega_f) \right],
\end{eqnarray}
with $h\ll  1$ the thickness  of the disk,  and $\Gamma\simeq 1$  is a
parameter that characterizes the  steepness in the vertical transition
from Keplerian rotation inside the disk to corotation with the star at
the top  of the disk.  In the derivation of  Equation~(\ref{eq:Tdint}), the
magnetic field  of the white dwarf  has been approximated as  a dipole
and it was assumed that the rotating axis of the star is perpendicular
to the  plane of the disk.  Note that this expression  generalizes the
models of \cite{1987A&A...183..257W,1995ApJ...449L.153W}.

Based on  this, the post-merger  evolution of the angular  momentum of
the white  dwarf is the  result of the  combined effect of  the dipole
radiation  torque, $T_{\rm  dipole}$,  the  accretion torque,  $T_{\rm
acc}$, the disk-interaction torque,  $T_{\rm disk}$, and the propeller
torque $T_{\rm prop}$:
\begin{equation}\label{eq:TotalTorque}
\dot{J}_{\rm WD}=\left\{ 
\begin{array}{lcl}
T_{\rm acc}& & R_{\rm WD}>R_{\rm 0}\\ 
& & \\
T_{\rm acc}+T_{\rm dipole}+T_{\rm mag} & & R_{\rm WD}<R_0 \,\wedge\, \omega_{\rm f}\leq 1\\
& & \\
T_{\rm dipole}+T_{\rm prop} & &R_{\rm WD}<R_0 \, \wedge\,\omega_{\rm f}> 1
\end{array}
\right.
\end{equation}
In  the simulations  presented here  we  assumed that  the inner  disk
radius is the Alfv\'en radius.

%%%%%%%%%%%%%%%%%%%%%%%%%%%%%%%%%%%%%%%%%%%%%%%%%
\subsection{Rotating white dwarf configurations}
\label{sec:WD}
%%%%%%%%%%%%%%%%%%%%%%%%%%%%%%%%%%%%%%%%%%%%%%%%%

We  follow  the evolution  of  the  spin  of the  post-merger  remnant
calculating at  every timestep  the new stable  rotating configuration
with  mass,  $M_{\rm  WD}+\delta  M$, and  angular  momentum,  $J_{\rm
WD}+\delta    J$,   adopting    the   slow    rotation   approximation
\citep{1968ApJ...153..807H,1967ApJ...150.1005H}. We  also assumed that
the central remnant product of the  merger rotates as a rigid body, as
predicted  by  detailed SPH  simulations  \citep{2009A&A...500.1193L}.
The equation of  state of the white  dwarf is assumed to be  that of a
zero temperature degenerate  electron gas \citep{1931ApJ....74...81C},
to which we add the corresponding contribution of ions.

The region of stability of uniformly rotating white dwarfs is bound by
the  secular  axisymmetric  instability limit,  the  mass-shedding  or
Keplerian limit, and the inverse $\beta$-decay instability limit, that
for   pure   carbon  is   $3.49\times   10^{10}\,   {\rm  g\,   cm^3}$
\citep{2011PhRvD..84h4007R, 2013ApJ...762..117B}.

\begin{figure}
\centering
\includegraphics[width=0.9\hsize,clip]{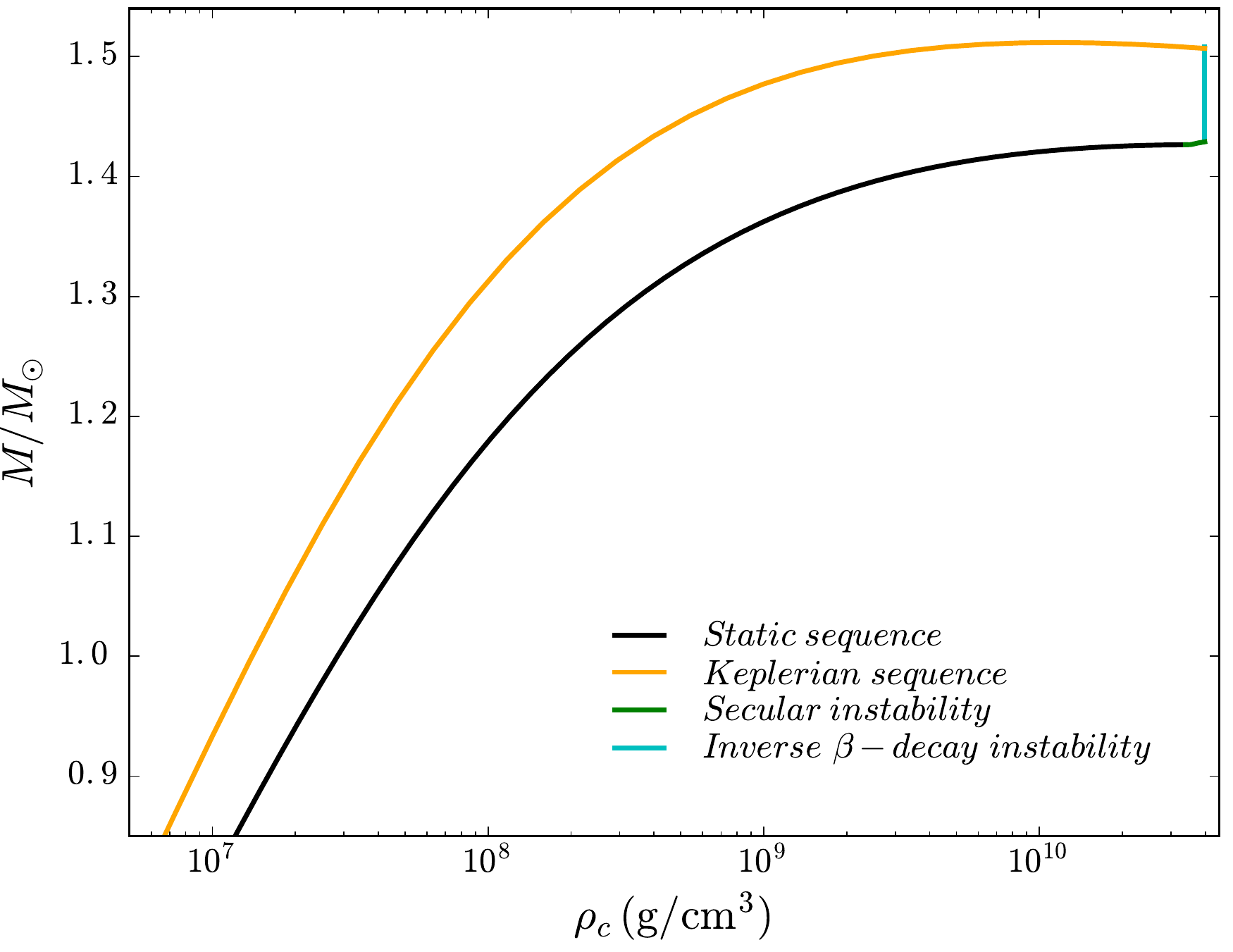}
\caption{Instability lines of rotating white  dwarfs in the diagram of
  mass versus  central density. The  stability region is  delimited by
  the  static sequence  (black line),  the Keplerian  sequence (orange
  line), the line of secular instability (green line), and the inverse
  $\beta$-decay line (blue).}
\label{fig:WDrot}
\end{figure}

Figure~\ref{fig:WDrot} shows  the mass-central density relation  for the
general  relativistic uniformly  rotating  white  dwarfs: the  static,
Keplerian, secular axisymmetric  instability and inverse $\beta$-decay
sequences  enclose  the  stability   region.   The  mass-shedding,  or
Keplerian,  limit is  reached when  the angular  velocity of  the star
equals the Keplerian  velocity of a particle orbiting  at the equator,
namely when the centrifugal and gravitational forces are balanced.  In
this situation matter  at the surface of the star  is marginally bound
to it.  Hence, small perturbations result  in mass loss until the star
becomes  either stable  again or  arrives to  a dynamical  instability
point  \citep{2000PhRvD..61d4012S,2003LRR.....6....3S}.  If  the white
dwarf  crosses  the  $\beta$-decay instability  limit,  electrons  are
captured by nuclei.  Since the  principal contribution to the pressure
of  the  star  comes  from electrons,  electron  captures  reduce  the
pressure, leading to  an instability \citep{1961ApJ...134..669S}.  The
secular axisymmetric  instability arises because the  star is unstable
with respect to axisymmetric perturbations. In a first phase, the star
is expected  to evolve quasi-stationarily with  the instability growth
timescale, which depends on the  time required for angular momentum to
be redistributed either by viscous  dissipation, or by the emission of
gravitational  waves \citep{1970ApJ...161..571C}.   This timescale  is
typically    long     compared    to    the     dynamical    timescale
\citep{2013ApJ...762..117B},  except in  the  non-rotating case  where
they are equal,  because in this case there is  no angular momentum to
be  redistributed.  Eventually,  when the  star crosses  the dynamical
instability    limit,     gravitational    collapse     takes    place
\citep{2003LRR.....6....3S}.    A  sufficient   (but  not   necessary)
condition for the  onset of secular instability can  be obtained using
the  turning-point  method  \cite{1988ApJ...325..722F}.   This  method
considers  a  sequence  of  uniformly   rotating  models  of  a  given
(constant)  angular momentum.   The  secular axisymmetric  instability
sets in at the density of the turning-point:
\begin{equation}
   \left( \frac{\partial\, M(\rho_c,J)}{\partial\, \rho_c} \right)_J=0.
   \label{eq:secinst}
\end{equation}
Finally,   the  maximum   mass   of  static   configurations  is   the
Chandrasekhar  limiting  mass  $M_{\rm  max}=M_{\rm  Ch}\approx  1.4\,
M_{\sun}$, while the maximum mass of rotating white dwarfs lies on the
Keplerian  sequence   $M_{\rm  max}^{J\neq   0}=1.52\,M_{\sun}$.   The
configurations with mass  between $M_{\rm max}^{J=0}<M_{\rm WD}<M_{\rm
max}^{J\neq 0}$  are called super-Chandrasekhar white  dwarfs, and are
metastable  since they  are  supported by  rotation.   As the  angular
momentum  of  the star  varies  the  central white  dwarf  necessarily
evolves  towards one  of the  previously described  instability limits
(mass-shedding or secular axisymmetric instability).

%%%%%%%%%%%%%%%%%%%%%%%%%%%%%%%%%%%%%%%%%%%%%%%%%
\subsection{Temperature of the central remnant}
\label{sec:TWD}
%%%%%%%%%%%%%%%%%%%%%%%%%%%%%%%%%%%%%%%%%%%%%%%%%

We turn now to the thermal evolution of the central white dwarf remnant. We have seen above in sections~\ref{sec:torque} and \ref{sec:WD}, and we shall see below from the energy balance and transport equations, that the structure and thermal evolution of the white dwarf depend crucially on the accretion rate, which is set by the disk physics. In section \ref{sec:mdotWD} we analyze the evolution and fate of the post-merger system for two physical prescriptions to set the accretion rate. In the first case the infall rate is driven by the disk angular-momentum transport (viscous) timescale which, owing to its shortness for the present binary parameters, leads to highly super-Eddington rates. It has been argued in the literature that the dissipation required to produce such very short viscous timescales might heat the disk to the point of carbon ignition \citep[see, e.g.,][]{1990A&A...236..378M}. Such heating effects are not accounted for in the disk model adopted here. In addition, we also compute the evolution of the system in the case when the matter infall is driven not by the viscous timescale but by the cooling timescale. We show that this assumption leads to accretion rates near the Eddington limit value.

The thermal evolution of an accreting  white dwarf has been studied by
many authors \citep{1985ApJ...297..531N,1985A&A...150L..21S, 2007MNRAS.380..933Y}.  The main goal of  most of these works  was to establish
whether  the conditions  that may  lead to  a type  Ia supernova  were
reached. For      instance,     in     a      pioneering     work
\citet{1985ApJ...297..531N} found that for accretion rates larger than
a   critical  value   $\dot{M}_{\rm  crit}\approx   2\times  10^{-6}\,
M_{\sun}\,{\rm yr}^{-1}$, the  outer layers of the star  are heated by
mass  accretion,  while heat  conduction  and  neutrino emission  cool
it. This results in a thermal  inversion near the surface of the star,
and   ultimately    leads   to   an   off-center    carbon   ignition.
\citet{1985A&A...150L..21S} followed  the subsequent evolution  of the
star and determined  that carbon is burned in the  entire white dwarf,
being the  final outcome an  oxygen-neon white dwarf.   However, these
works  only   considered  non-rotating  configurations   and  constant
accretion rates.  More recently,  \citet{2007MNRAS.380..933Y} studied
the viscous evolution  of remnants of white  dwarf mergers introducing
the  effects of  rotation  but did  not consider  the  effects of  the
magnetic field of the white dwarf.  The post-merger object was modeled
as a  differentially rotating white dwarf  with a cold core  and a hot
envelope  accreting  mass  from  a surrounding  Keplerian  disk  at  a
constant  rate.  This  allowed them  to compute  the conditions  under
which the off-center ignition can be avoided for the case in which the
magnetic field is not taken into account   

We have explained above (see Sec.~\ref{sec:torque}) how the rotation  rate of the  merged remnant  evolves due  to the  interaction of  the magnetic
field and  the disk, and  which are the equilibrium  configurations of
the  the rotating  white dwarf.   However,  during the  course of  the
evolution  the temperature  of the  white  dwarf also  changes due  to
accretion from the  disk, which heats the white  dwarf interior. This,
in  turn, may  trigger a  type Ia  supernova explosion.   In order  to
assess  this possibility,  we  follow the  evolution  of the  interior
temperature of the  post-merger central white dwarf  in an approximate
way, which is described next.

The equation of energy conservation reads:
\begin{equation}
	\frac{dL}{dm}=\epsilon_{\rm nuc}-\epsilon_\nu+T\dot{s},
	\label{eq:energy_conser}
\end{equation}
where $m$ is  the mass coordinate (i.e., the mass  enclosed within the
radial distance $r$), $L$ is  the luminosity, $T$ the temperature, $s$
the specific entropy, and  $\epsilon_{\rm nuc}$ and $\epsilon_\nu$ are
the energy release and energy loss  per unit mass by nuclear reactions
and neutrino emission, respectively.

For carbon-oxygen  white dwarfs,  thermonuclear energy  is essentially
released by two nuclear reactions:
\begin{eqnarray}
^{12}{\rm C}+^{12}{\rm C}\,&\rightarrow&\,^{20}{\rm Ne}+\alpha+4.62\,{\rm MeV},\\
^{12}{\rm C}+^{12}{\rm C}\,&\rightarrow&\,^{23}{\rm Na}+{\rm p}+2.24\,{\rm MeV},
\end{eqnarray}
with  nearly  the  same  probability. We  adopted  the  carbon  fusion
reaction rates of \citet{2005PhRvC..72b5806G}, which are valid for all
regimes of  $\rho$ and $T$, that  is from the thermonuclear  regime to
the pynonuclear regime.   For the neutrino energy losses,  we used the
analytical   fits  of   \citet{1996ApJS..102..411I},  which   consider
electron-positron                   pair                  annihilation
($e^{-}e^{+}\rightarrow\nu\bar{\nu}$),     photo-neutrino     emission
($e+\gamma\rightarrow       e\nu\bar{\nu}$),       plasmon       decay
($\gamma\rightarrow\nu\bar{\nu}$), and electron-nucleus bremsstrahlung
[$e(Z,A)\rightarrow e(Z,A)\nu\bar{\nu}$].  As it  will be discussed in
below, the dominant channel for neutrino losses in the central regions
is  electron-nucleus  bremsstrahlung,  while   in  the  outer  layers,
emission of neutrinos is dominated by plasmon decay.

The energy flux is given by:
\begin{equation}
\frac{dT}{dr}=-\frac{3}{16\sigma}\frac{\kappa \rho}{T^3}\frac{L}{4\pi r^2},
\label{eq:gradT}
\end{equation}
where $\sigma$ is  the Stephan-Boltzmann constant and  $\kappa$ is the
opacity, which can be written as:
\begin{equation}
\frac{1}{\kappa}=\frac{1}{\kappa_{\rm cond}}+\frac{1}{\kappa_{\rm rad}},
\label{eq:opacity}
\end{equation}
with  $\kappa_{\rm rad}$  and $\kappa_{\rm  cond}$ the  Rosseland mean
radiative opacity  and the  conductive opacity, respectively.   In the
white  dwarf  core degeneracy  is  so  high  that the  most  efficient
transport mechanism is conduction. Hence,  the opacity is dominated by
the   first   term.    We   adopted  the   thermal   conductivity   of
\citet{1983ApJ...273..774I,1984ApJ...285..758I},   whereas   for   the
radiative opacity  we used Kramer's law,  $\kappa_{\rm rad}=4.34\times
10^{24}\rho T^{-3.5}$~cm$^2$~g$^{-1}$.

The change of entropy with time can be obtained from:
\begin{equation}
  T\dot{s}=c_v \dot{T}-\left[ \frac{P}{\rho^2}-
  \left(\frac{\partial\, u}{\partial\, \rho} \right)_T \right]\dot{\rho}
\label{eq:entropy_evol}
\end{equation}
where $u$ is the specific internal  energy and $c_v$ the specific heat
capacity at constant volume.  The first term of the right-hand side of
Equation~(\ref{eq:entropy_evol}) corresponds to the release of the internal
energy, while the second term accounts for the change of gravitational
potential  energy  due   to  the  expansion  or   compression  of  the
configuration.

To evaluate the term in square brackets in Equation~(\ref{eq:entropy_evol})
we considered  a fully  ionized non-ideal electron-ion  plasma, taking
into account  the ion-ion  and the ion-electron  Coulomb interactions,
and also  the exchange-correlation corrections of  electrons.  We note
that for super-Chandrasekhar white dwarfs both Coulomb corrections and
quantum effects are important.   The importance of Coulomb corrections
is measured  by the  Coulomb coupling  parameter $\Gamma=Z^2e^2/(aT)$,
where  $a=(3/(4\pi n_i)^{1/3}$  is the  mean interaction  distance and
$n_i$ the ion  number density.  At $\Gamma\lesssim 1$  the ions behave
as a  gas, at $\Gamma>1$ as  a strongly coupled Coulomb  liquid, while
crystallization occurs at $\Gamma\approx 175$.  Quantum effects become
important  at temperatures  smaller  than $T_{\rm  p}=\hbar\omega_{\rm
p}/k_{\rm  B}$, where  $\omega_{\rm p}=(4\pi  Z^2e^2n_i/m_i)^{1/2}$ is
the ion  plasma frequency.   Finally, we used  analytical fits  to the
heat   capacity   obtained   from   the  free   energy   computed   by
\citet{1998PhRvE..58.4941C} and \citet{2000PhRvE..62.8554P}.

Following \citet{1982ApJ...253..798N}, we  obtained the density change
by  adopting  the  white  dwarf  mass  coordinate,  $q_{\rm  WD}\equiv
m/M_{\rm WD}$, as the independent variable:
\begin{equation}
  \dot{\rho}=\left[ \left( \frac{\partial \, \rho}
    {\partial M_{\rm WD}} \right)_{q_{\rm WD}} -\frac{q_{\rm WD}}{M_{\rm WD}}\left( \frac{\partial \, \rho}
    {\partial q_{\rm WD}} \right)_{M_{\rm WD}} \right]\dot{M}_{\rm WD}
\label{eq:rho_evol}
\end{equation}
This  equation  explicitly provides  the  contribution  of the  global
structural  changes  as  well  as the  contribution  of  the  internal
distribution of density to the  thermal evolution during the accretion
process.

The post-merger evolution of the  system is computed assuming that the
central white  dwarf evolves  in a  sequence of  stable configurations
(see   Section\ref{sec:WD}).   To   calculate  the   evolution  of   the
temperature,      at     each      time     step      we     integrate
Equations~(\ref{eq:energy_conser})      and     (\ref{eq:gradT})      using
Equations~(\ref{eq:entropy_evol}) and (\ref{eq:rho_evol}).

In    order    to    integrate    Equations~(\ref{eq:energy_conser})    and
(\ref{eq:gradT}), a set  of boundary conditions at the  surface of the
star must be adopted. We treat  the material accreted in each interval
time   as   a  thin   envelope   surrounding   the  star.    Following
\citet{2004ApJ...600..390T},  we  assume  that the  accreted  material
enters  on top  of the  shell and  pushes down  the existing  material
deeper into the star. Then, the local heat equation becomes:
\begin{equation}
  T\left(\frac{\partial}{\partial t} + v_r\frac{\partial}{\partial r}\right)
  s\approx T v_r\frac{\partial s}{\partial r}=\frac{dL}{dm}-
  (\epsilon_{\rm nuc}-\epsilon_{\nu})
\label{eq:DT_envelope}
\end{equation}
where  $v_r=\dot{M}_{\rm WD}/(4\pi  r^2\rho)$ is  the velocity  of the
material given  by mass conservation. We  constructed static envelopes
for each stable configuration with  total mass $M_{\rm WD}$ and radius
$R_{\rm      WD}$,      integrating     Equations~(\ref{eq:gradT})      and
(\ref{eq:DT_envelope}).

To  analyze  if  the  newly  formed white  dwarf  reaches  during  its
evolution the conditions suitable to produce  a type Ia we adopted the
following procedure.  We  first require as a  necessary condition that
the white  dwarf crosses the  ignition curve,  i.e.  the curve  in the
$\log\rho-\log  T$ plane  where  the nuclear  energy released  becomes
larger   than  the   neutrino   emissivity,   $\epsilon_{\rm  nuc}   =
\epsilon_{\nu}$.   For temperatures  and densities  beyond this  curve
nuclear  energy production  exceeds neutrino  losses and  the star  is
heated, possibly  leading to  a supernova outburst.   Additionally, we
require that  burning proceeds  in an  almost instantaneous  way.  The
characteristic time $\tau_{\rm CC}$ of nuclear reactions is:
\begin{equation}
  \tau_{\rm CC} = \frac{\epsilon_{\rm nuc}}{\dot{\epsilon}_{\rm nuc}}
  \approx \epsilon_{\rm nuc}\biggl(\dot{T}
  \frac{\partial\epsilon_{\rm nuc}}{\partial T} \biggr)^{-1} =
  c_p\biggl(\frac{\partial\epsilon_{\rm nuc}}{\partial T} \biggr)^{-1},
    \label{eqn:instant-ch-time}
\end{equation}
where  $c_p$  is the  specific  heat  at  constant pressure.  If  this
characteristic timescale becomes shorter  that the dynamical timescale
\begin{equation}
  \tau_{\rm dyn}^{-1}=\sqrt{24\pi  G\rho},
\end{equation}
the star reaches the thermodynamic  conditions necessary to explode as
a type Ia supernova.

%%%%%%%%%%%%%%%%%%%%%%%%%%%%%%%%%%%%%%%%%%%%%%%%%
%%%%%%%%%%%%%%%%%%%%%%%%%%%%%%%%%%%%%%%%%%%%%%%%%
\section{Mass accretion rate on the central remnant}
\label{sec:mdotWD}
%%%%%%%%%%%%%%%%%%%%%%%%%%%%%%%%%%%%%%%%%%%%%%%%%
%%%%%%%%%%%%%%%%%%%%%%%%%%%%%%%%%%%%%%%%%%%%%%%%%

As mentioned  before, the accretion rate  onto the post-merger
  central remnant  can be computed  in two different ways.  Within the
  first approximation, the accretion rate  is set by the thermal state
  of       the      envelope       of       the      white       dwarf
  \citep{2007MNRAS.380..933Y}.  Within  the  second  one  the  central
  remnant accretes mass  from a thin Keplerian disk  and its evolution
  is  given by  the viscous  time-scale  -- see  \cite{vk10} but  also
  \cite{Ji}.  Since,  due to  the lack of  full numerical  analyses of
  this  process with  realistic physical  ingredients, it  is not  yet
  clear which  one of  these treatments is  more appropriate,  we will
  calculate the evolution of the  central white dwarf remnant adopting
  both prescriptions and we will  discuss the differences between both
  sets of calculations. In both cases we are interested in determining
  the long-term evolution central white  dwarf taking into account the
  interaction between the  disk and the magnetic  field resulting from
  the merger.  To  handle this problem, we adopt  a simplified picture
  of the post-merger system consisting  of a magnetized rotating white
  dwarf surrounded by a thin Keplerian disk.

%%%%%%%%%%%%%%%%%%%%%%%%%%%%%%%%%%%%%%%%%%%%%%%%%
\subsection{Accretion rate set by the thermal timescale}
\label{sec:mdot_envelope}
%%%%%%%%%%%%%%%%%%%%%%%%%%%%%%%%%%%%%%%%%%%%%%%%%

As  discussed by  \citet{2007MNRAS.380..933Y}, the  assumption
  that the accretion timescale is  set by the viscous time-scale given
  by the Shakura-Sunyaev $\alpha$  prescription could be inappropriate
  to  estimate the  accretion  timescale onto  the newly-formed  white
  dwarf  since the  structure of  the  merged system  deviates from  a
  central,  point-like mass  surrounded by  a  thin disk  that is  the
  central   assumption   adopted    within   the   $\alpha$-formalism.
  \citet{2007MNRAS.380..933Y} argued  that under these  conditions the
  relevant timescale might be the cooling timescale of the low density
  envelope between the  central object and the disk.   If the relevant
  timescale  is  the   cooling  timescale  of  the   envelope  of  the
  post-merger white dwarf, the accretion rate will be given by:

\begin{equation}
\dot{M}_{\rm WD}=\frac{M_{\rm disk}}{{\rm min}(\tau_{\nu},\tau_{\rm thermal})}
\label{eq:MdotWD}
\end{equation}

where $M_{\rm disk}$ is the total mass of the disk, and $\tau_\nu$ and $\tau_{\rm thermal}$ are the neutrino cooling time and the thermal time on the envelope of the white dwarf, respectively. These two  timescales are given by:
\begin{equation}\label{eq:tau_neutrino}
\tau_{\nu}=\left.\frac{c_{\rm v}}{\epsilon_{\nu}}{T_S}\right|_S,
\end{equation}
and 
\begin{equation}
\tau_{\rm thermal}=\frac{3}{64\sigma} \left(\int_{\Delta r}\left( \frac{\kappa c_v}{T^3}\right)^{1/2}\rho\, dr\right)^2,
\end{equation}
\citep{1969ApJ...156..549H, 2003ApJ...583..885P},  where $c_v$  is the
heat   capacity  at   constant  volume,   $\kappa$  is   the  opacity,
$\epsilon_\nu$  is the  energy release  by the  neutrino emission  and
$\Delta r$ delimits the region  of interest -- see Section~\ref{sec:TWD}
for a discussion of the thermal properties of the white dwarf.

For  the  values  typical   of  the  post-merger  white  dwarf
  ($\rho=10^6$~g~cm$^{-3}$ and  $T=10^8$~K),  the neutrino
  luminosity  will  be $L_\nu\sim  10^2\,  L_{\sun}$  and the  thermal
  energy will be  of $U\sim 10^{48}$~erg. At  the beginning of
  the   simulations   the   neutrino  cooling   timescale   is   about
  $\tau_\nu\sim  6\times  10^5$~yr  while  the  thermal  timescale  is
  $\tau_{\rm thermal}\sim  10^6$~yr. Then, the initial  accretion rate
  is about $\dot{M}_{\rm  WD}\sim 10^{-7}\, M_{\sun}$~yr$^{-1}$, close
  to the Eddington limit.

\begin{figure}
\centering
\includegraphics[width=0.9\hsize,clip]{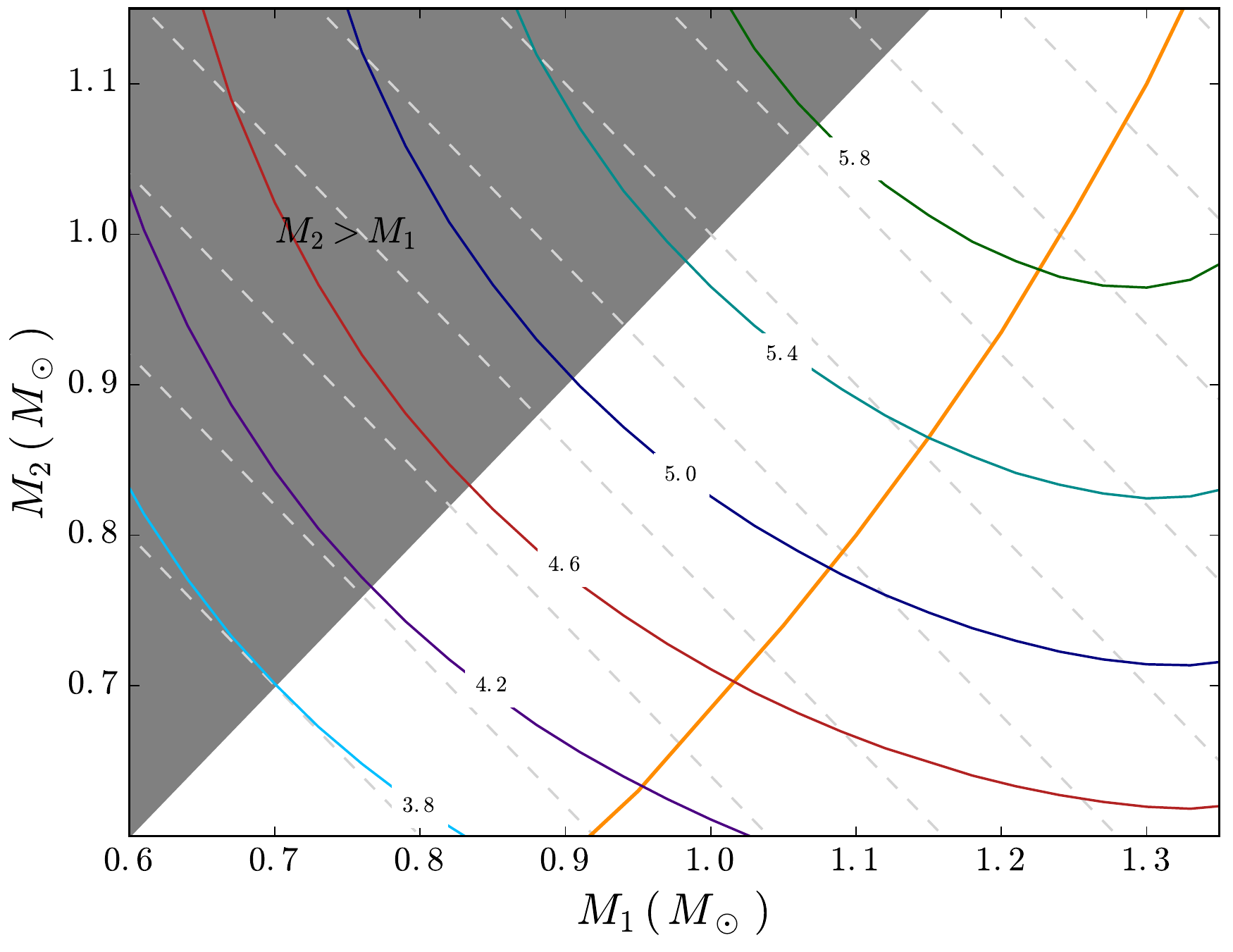}
\caption{Parameter   space  of   the   initial   white  dwarf   binary
  system. Solid lines are contours of constant total angular momentum,
  labeled in  units of  $10^{50}$~g~cm$^2$~s$^{-1}$. The  gray, dotted
  straight  lines are  contours of  constant total  mass, from  $1.4\,
  M_{\sun}$ to  $2.6\, M_{\sun}$,  in steps  of $0.1\,  M_{\sun}$. The
  shaded area corresponds to configuration with $q>1$. The orange line
  corresponds to location of the  system when the secondary is filling
  its Roche lobe ($R_{2}=R_{\rm L}$).}
\label{fig:M1M2}
\end{figure}

%%%%%%%%%%%%%%%%%%%%%%%%%%%%%%%%%%%%%%%%%%%%%%%%%
\subsection{Accretion rate set by the viscous timescale}
\label{sec:disk}
%%%%%%%%%%%%%%%%%%%%%%%%%%%%%%%%%%%%%%%%%%%%%%%%%

Numerical  simulations   --  see,   e.g.,  \cite{1990ApJ...348..647B},
\citet{2009A&A...500.1193L}  and  \cite{2011ApJ...737...89D}  --  show
that  the disk  product  of the  coalescence of  two  white dwarfs  of
unequal  masses  is  not  thin,  although  it  is  not  thick  either.
Specifically, \citet{2009A&A...500.1193L} found  that the thickness of
the newly formed disk is $H\sim 5.0\times 10^{-3}~R_{\sun}$, while the
typical size is $R_{\rm disk}\sim  0.2~R_{\sun}$ -- see their Table~1.
Hence, $H/R_{\rm disk}\simeq  0.025$. Thus, assuming that  the disk is
thin is not an extreme assumption and, in the worst of the situations,
it  can  be considered  as  a  limiting  case.  Obviously,  the  other
limiting case  is to assume  that the disk  is thick.  Here  we assume
that the disk is thin, and we postpone  the study of a thick disk to a
forthcoming  publication.

For a  thin, Keplerian  accretion disk, the  evolution of  the surface
mass density, $\Sigma$, is given by the diffusion equation:
\begin{equation}\label{eq:DiskSurDensity_evol}
  \frac{\partial\, \Sigma}{\partial t}=\frac{3}{r}\frac{\partial}{\partial r}
  \left[r^{1/2}\frac{\partial }{\partial r}
  \left(\nu \Sigma r^{1/2}\right)\right],
\end{equation}
where $\nu$ is the turbulent kinematic viscosity. To describe the time
evolution of the  disk we use one of the  three self-similar solutions
of         Equation~(\ref{eq:DiskSurDensity_evol})         found         by
\citet{1974PhDT.......131P}. These solutions are obtained  for an opacity parametrized as $\kappa=\kappa_0\rho^aT^b$.
Within   this  approximation   mass  is
accreted onto the central white dwarf, but the angular momentum of the
disk is conserved, $J_{\rm disk}= J_0$,  because the outer edge of the
disk moves outwards -- see Equation~(\ref{eq:diskRout}).  In a general case $J_{\rm disk}$ would not be conserved  because there is a trade-off of angular moment between the disk and the central white dwarf.  However,
since the  angular moment of  the white  dwarf, $J_{\rm WD}$,  is much
smaller than  that of disk, this  solution is accurate enough  for the
purpose  of  estimating  the  accretion rate  onto  the  white  dwarf.
Actually,  we  have  checked  that at  all  times  $J_{\rm  WD}/J_{\rm
disk}<0.1$ in our simulations.

Within this approximation  mass flows at the inner disk  boundary at a
rate
\begin{equation}\label{eq:Mdisk}
  \dot{M}_{\rm disk}=\frac{(\beta-1)M_0}{t_{\rm visc}}
  \left(1+\frac{t}{t_{\rm visc}}\right)^{-\beta},
\end{equation}
and the disk outer radius evolves as
\begin{equation}
  R_{\rm out}=r_0\left( 1+\frac{t}{t_{\rm visc}} \right)^{2\beta},
  \label{eq:diskRout}
\end{equation}
where $M_0$ and $J_0$ are,  respectively, the initial mass and angular
momentum  of the  disk,  $j_0=J_0/M_0$, $r_0=3.88\,j_0^2/(GM_0)$,  and
$\beta$ is  a constant that  depends on  the opacity regime.   Here we
have  adopted a  bound-free  opacity ($\beta=5/4$). The  viscous  timescale   within this model is
determined by $M_0$ and $j_0$ as follows \citep{2009ApJ...702.1309E}:
 \begin{equation}\label{eq:tvisc}
 t_{\rm visc}=9.82\times 10^{9} \frac{j_0^{25/7}}{M_{\rm WD}^{10/7}M_0^{3/7}}\left(\frac{\sigma}{\alpha^8\kappa_0}\right)^{1/7}\left(\frac{\bar{\mu}m_p}{\kappa_B}\right)^{15/14}
\end{equation}
where $m_p$ is the proton mass and $\mu$ the mean molecular weight. Adopting $\alpha=0.1$  for the viscosity parameter \citep{SS73} gives: 
\begin{equation}
  t_{\rm visc}\simeq 10.9\,\left(\frac{j_0}{10^{18}\,{\rm cm^2\,s^{-1}}}\right)
  ^{25/7}\left( \frac{M_0}{0.1\, M_{\sun}} \right)^{-3/7} \, {\rm s},
\label{eq:ViscousScale}
\end{equation}
This  solution  has been  employed  to  describe debris  disks  around
massive   black   holes   formed   by  tidal   disruption   of   stars
\citep{1990ApJ...351...38C}  and in  supernova  fallback disks  around
young   neutron    stars   \citep{2000ApJ...534..373C}.  This
  formulation was also  used in \citet{kulebi} to  study the long-term
  evolution  of  the  disk   interaction  of  magnetized  white  dwarf
  resulting  from  white dwarf  mergers  that  do not  develop  prompt
  explosion conditions.

We obtain that the initial conditions of the merged system are
  such      that      $t_{\rm      visc}\sim      0.2-0.8$~s      (see
  Table~\ref{tab:InConModels}).  This,  in turn, results  in accretion
  rates  at  early   times  $\dot{M}_{\rm  WD}=M_{0}/t_{\rm  visc}\sim
  10^{-1}\, M_{\sun}$~s$^{-1}$  and initial disk  outer-radius $R_{\rm
    out}\sim  0.1R_{\sun}$. However,  for  times  longer than  $t_{\rm
    visc}$, the accretion rate drops one order of magnitude and, after
  $10^2$~s, the accretion is $10^{-4}\, M_{\sun}$~s$^{-1}$.

In the above equations the disk viscous timescale, $t_{\rm visc}$, is a constant. It is clear from Equation.~(\ref{eq:tvisc}) that it is a good approximation for low infall rates because the specific angular momentum would not change appreciably. The above estimate of the accretion rate shows that this is not the present case so we have checked the effect of the assumption of the constancy of $t_{\rm visc}$ on the evaluation of $\dot{M}_{\rm disk}$ as follows. First we compute the accretion rate using Equation~(\ref{eq:Mdisk}) keeping $t_{\rm visc}$ constant and then allowing it to increase with time owing to the mass loss by the disk. We found that the increase of $t_{\rm visc}$ would lower $\dot{M}_{\rm disk}$ at most by 10\% during the evolution. This result assure us about our assumption of a constancy of the viscous timescale on the estimate of the infall rate at the inner disk radius.

%%%%%%%%%%%%%%%%%%%%%%%%%%%%%%%%%%%%%%%%%%%%%%%%%
\section{Initial conditions}
\label{sec:initial}
%%%%%%%%%%%%%%%%%%%%%%%%%%%%%%%%%%%%%%%%%%%%%%%%%

\begin{table*}
\centering
\caption{Parameters of the simulations of post-merger remnants.}
\begin{tabular}{cccccccccccccc}
  \hline
  \hline
  Model & $M_{1}$ & $M_2$ & $M_{\rm WD}$ &$M_{0}$ &$R_{\rm eq,WD}$ &$B_{\rm WD}$  & $\Omega_{\rm WD}$ & $J_{\rm WD}$&    $J_0$ & $\varepsilon$ & $t_{\rm dyn}$ & $t_{\rm visc}$ & $t_{\rm cool}$  \\ 
  & $[M_{\sun}]$ & $[M_\sun]$ & $[M_\sun]$ & $[M_\sun]$ & $[R_\sun]$ &[G] &  $[{\rm rad\, s^{-1}}]$ &$[{\rm g\, cm^2\,s^{-1}}]$ &  $[{\rm g\, cm^2\,s^{-1}}]$ &$[-]$ & $[{\rm s}]$ &$[{\rm s}]$&$[{\rm yr}]$\\
  \tableline
  A & $1.12$ & $0.78$ & $1.45$ & $0.45$ & $0.0026$ & $10^9$ & $3.00$ & $2.79\times 10^{49}$ & $5.86\times 10^{50}$ & 0.1 & 0.3 &$0.745$ & $2.26\times 10^4$ \\
  B & $1.12$ & $0.78$ & $1.45$ & $0.45$ & $0.0026$ & $10^6$ & $3.00$ & $2.79\times 10^{49}$ & $5.86\times 10^{50}$ & 0.1 & 0.3 & $0.745$ & $2.26\times 10^4$ \\
  C & $1.12$ & $0.78$ & $1.45$ & $0.45$ & $0.0026$ & $10^7$ & $3.00$ & $2.79\times 10^{49}$ & $5.86\times 10^{50}$ & 0.1 & 0.3 & $0.745$ & $2.26\times 10^4$ \\
  D & $1.12$ & $0.78$ & $1.45$ & $0.45$ & $0.0026$ & $10^8$ & $3.00$ & $2.79\times 10^{49}$ & $5.86\times 10^{50}$ & 0.1 & 0.3 & $0.745$ & $2.26\times 10^4$ \\
\hline
  E & $1.12$ & $0.78$ & $1.45$ & $0.45$ & $0.0026$ & $10^9$ & $3.00$ & $2.79\times 10^{49}$ & $5.86\times 10^{50}$ & 0.1 & 0.3 & $0.745$ & $2.26\times 10^4$ \\
  F & $1.12$ & $0.78$ & $1.45$ & $0.65$ & $0.0026$ & $10^9$ & $3.00$ & $2.79\times 10^{49}$ & $5.86\times 10^{50}$ & 0.1 & 0.3 & $0.253$ & $2.26\times 10^4$ \\
  G & $1.12$ & $0.78$ & $1.45$ & $0.45$ & $0.0026$ & $10^6$ & $3.00$ & $2.79\times 10^{49}$ & $5.86\times 10^{50}$ & 0.1 & 0.3 & $0.745$ & $2.26\times 10^4$ \\
  H & $1.12$ & $0.78$ & $1.45$ & $0.45$ & $0.0026$ & $10^9$ & $3.00$ & $2.79\times 10^{49}$ & $5.86\times 10^{50}$ & 0.5 & 0.3 & $0.745$ & $2.26\times 10^4$ \\
  I & $1.12$ & $0.78$ & $1.45$ & $0.45$ & $0.0039$ & $10^9$ & $2.30$ & $4.56\times 10^{49}$ & $5.69\times 10^{50}$ & 0.1 & 0.4 & $0.668$ & $2.79\times 10^3$ \\
  J & $1.12$ & $0.78$ & $1.45$ & $0.45$ & $0.0034$ & $10^9$ & $2.50$ & $3.87\times 10^{49}$ & $5.76\times 10^{50}$ & 0.1 & 0.4 & $0.697$ & $5.66\times 10^3$ \\
  K & $1.12$ & $0.78$ & $1.45$ & $0.45$ & $0.0032$ & $10^9$ & $2.60$ & $3.60\times 10^{49}$ & $5.78\times 10^{50}$ & 0.1 & 0.4 & $0.709$ & $7.79\times 10^3$ \\
  L & $1.12$ & $0.78$ & $1.45$ & $0.45$ & $0.0030$ & $10^9$ & $2.70$ & $3.35\times 10^{49}$ & $5.83\times 10^{50}$ & 0.1 & 0.4 & $0.719$ & $1.05\times 10^4$ \\
  M & $1.12$ & $0.78$ & $1.45$ & $0.45$ & $0.0029$ & $10^9$ & $2.80$ & $3.14\times 10^{49}$ & $5.68\times 10^{50}$ & 0.1 & 0.4 & $0.729$ & $1.38\times 10^4$ \\
  N & $1.12$ & $0.78$ & $1.45$ & $0.45$ & $0.0028$ & $10^9$ & $2.85$ & $2.93\times 10^{49}$ & $5.68\times 10^{50}$ & 0.1 & 0.4 & $0.665$ & $1.57\times 10^4$ \\
  O & $1.12$ & $0.78$ & $1.45$ & $0.45$ & $0.0028$ & $10^9$ & $2.90$ & $2.93\times 10^{49}$ & $5.68\times 10^{50}$ & 0.1 & 0.3 & $0.737$ & $1.78\times 10^4$ \\ 
  P & $1.12$ & $0.78$ & $1.45$ & $0.45$ & $0.0022$ & $10^9$ & $3.50$ & $2.24\times 10^{49}$ & $5.86\times 10^{50}$ & 0.1 & 0.3 & $0.769$ & $5.34\times 10^4$ \\
  Q & $1.12$ & $0.78$ & $1.45$ & $0.45$ & $0.0017$ & $10^9$ & $4.50$ & $1.75\times 10^{49}$ & $5.96\times 10^{50}$ & 0.1 & 0.2 &$0.793$ & $8.79\times 10^4$ \\
\hline
\hline
\end{tabular}
\label{tab:InConModels}
\end{table*}

We compute  the post-merger  initial configuration assuming  that both
mass and angular momentum are conserved  during the merger.  This is a
reasonable  assumption, since  SPH simulations  show that  very little
mass  is ejected  from the  system.  Moreover,  little angular
  momentum is carried away by the unbound material, since its velocity
  is mostly radial.  A rough estimate  of the degree to which mass and
  angular momentum  are conserved is  $J_{\rm ej}/J \sim  M_{\rm ej}/M
  \sim   10^{-3}$  \citep{2009A&A...500.1193L,   2014MNRAS.438...14D}.
  Consequently, the  orbital angular momentum of  the coalescing white
  dwarfs is invested in spinning up the primary white dwarf and in the
  angular momentum of the rapidly rotating disk.

If the spin angular momentum of the merging white dwarfs is neglected,
the total angular momentum just before the merger is:
\begin{equation}
J_{\rm sys}=q\sqrt{\frac{GM_2^3(R_1+R_2)}{(1+q)}}
\label{eq:InitialCondition}
\end{equation}
where $(M_1,R_1)$  and $(M_2,R_2)$  are, respectively, the  masses and
radii of the  merging stars, and $q=M_2/M_1$ is the  mass ratio of the
original  binary system.   Figure~\ref{fig:M1M2} shows  the contours  of
constant  angular momentum  and of  constant total  mass in  the plane
defined by  $M_1$ and $M_2$. Given  an initial total mass  and angular
momentum of  the remnant,  ($M_{\rm WD},J_{\rm  WD})$, and  an initial
disk  mass, $M_0$,  we computed  the initial  angular momentum  of the
disk,  $J_0$, assuming  that the  central remnant  rotates as  a rigid
body. To  do this  we considered  that just  before the  mass transfer
episode the orbital  separation was such that the  secondary was about
to  fill its  Roche  lobe,  for which  we  adopted  the expression  of
\cite{1983ApJ...268..368E},  which  is  in  all  the  cases  close  to
$R_1+R_2$. This is the same to say that the merger episode begins when
both stars are  in contact.  We also took into  account that according
to detailed SPH simulations of the  merger process roughly half of the
mass of the secondary star goes to  form the disk, whereas the rest of
the material  is directly accreted  onto the primary component  of the
binary system \citep{2009A&A...500.1193L}.

Nevertheless,  given   the  exploratory  nature  of   the  simulations
presented here, we did not  constrain ourselves to the values obtained
in the SPH  simulations, and we adopted different values  for the mass
of the remnant and for the mass of the disk.  Since a rigidly rotating
white   dwarf   can   support   a  maximum   angular   momentum   $\la
10^{50}$~g~cm$^2$~s~$^{-1}$    \citep{2013ApJ...762..117B},    angular
momentum  conservation  implies that  a  substantial  fraction of  the
angular momentum of the binary system goes to the Keplerian disk.

Also,   we   mention   that  detailed   SPH   simulations   of
  non-symmetric  white dwarf  mergers show  that as  the less  massive
  white dwarf is disrupted and part of its matter is accreted onto the
  primary star, the accreted mass is compressed and heated.  Thus, for
  all  the cases  studied  here  we have  assumed  an inverse  initial
  temperature profile.   Specifically, we adopted  initial temperature
  profiles similar to those resulting from detailed SPH simulations.

In Table~\ref{tab:InConModels} we list the initial masses of the stars
of the original  binary system -- columns 2 and  3, respectively -- as
well  as the  initial  conditions  for the  models  presented in  this
work. The top section of  this table lists the characteristics
  of the  models for which  the accretion rate was  computed employing
  the cooling timescale, while the  bottom section summarizes the most
  relevant  information   of  those  models  computed   employing  the
  accretion rate given by the viscous timescale. Note that the mass of
  the rapidly rotating  central white dwarf $M_{\rm WD}$  of all these
  simulations (listed in column 4) is the same.  Since we are modeling
  white dwarfs as rigidly rotating configurations, only two parameters
  are needed to determine the  remaining characteristics of the merged
  configuration.   Thus,  we  decided  to  vary  the  initial  angular
  velocity $\Omega_{\rm WD}$ and the mass of the disk $M_0$. These two
  quantities  are listed  in columns  8  and 5,  respectively.  For  a
  super-Chandrasekhar  white dwarf  of  $1.45\,M_{\sun}$, the  minimum
  angular velocity for the gravitational stability to set in is around
  $2.03$~rad~s$^{-1}$. Note  that the angular velocities  adopted here
  are in all  cases larger than this value. Once  the angular velocity
  of the white dwarf is known we determine its total angular momentum,
  $J_{\rm WD}$,  which is  listed in  column 9 of  the table,  and the equatorial
  radius of the white dwarf, $R_{\rm WD}$, listed in the sixth column.
  The adopted magnetic field $B_{\rm WD}$  of the white dwarf is given
  in column 7. The rest of columns of Table~\ref{tab:InConModels} list
  the  initial angular  momentum $J_0$  (column 10)  of the  disk, the
  efficient parameter $\varepsilon$ (column  11), that is relevant for
  estimating the  mass accretion  rate,  the dynamical timescale $t_{\rm dyn}$, the viscous  timescale $t_{\rm
    visc}$,  and the  cooling  timescale of  the  disk $t_{\rm  cool}$
  (columns 12, 13 and 14).

In our simulations the magnetic  field of the remnant was kept
  fixed at $B_{\rm WD}=10^9$~G, a representative value of the magnetic
  field  resulting  from  the  stellar dynamo  originated  during  the
  coalescence  \citep{2012ApJ...749...25G}.   For the  typical  values
  found in  the SPH simulations  of merging white dwarfs  the magnetic
  field can be as high as $\sim 10^{10}$~G, depending if the dynamo is
  saturated. However,  we decided to  adopt a conservative value  as a
  reference magnetic field, $10^9$~G.   However, we also computed some
  models with various magnetic field strengths, ranging from $10^6$ to
  $10^9$~G.  An important point which is worth emphasizing is that for
  the simulations  presented here we  assumed that during  the merging
  process an ordered  global dipole field is  produced.  However, this
  is  not guaranteed  --  see,  for instance,  the  discussion in  the
  conclusions of \cite{2012MNRAS.427..190S}.

%%%%%%%%%%%%%%%%%%%%%%%%%%%%%%%%%%%%%%%%%%%%%%%%%
%%%%%%%%%%%%%%%%%%%%%%%%%%%%%%%%%%%%%%%%%%%%%%%%%
\section{Results}
\label{sec:results}
%%%%%%%%%%%%%%%%%%%%%%%%%%%%%%%%%%%%%%%%%%%%%%%%%
%%%%%%%%%%%%%%%%%%%%%%%%%%%%%%%%%%%%%%%%%%%%%%%%%

Since  the   evolution  of  the  post-merger   system  depends
  critically on the  adopted prescription for the  mass accretion rate
  onto the central  white dwarf resulting from the  coalescence of the
  binary system, we  discuss the results obtained  using the accretion
  rates described in Sections~\ref{sec:disk} and \ref{sec:mdot_envelope}
  separately.   We  first  discuss   the  results  obtained  using  an
  accretion rate  set by  the cooling  timescale, and  subsequently we
  present the results obtained when the viscous timescale is employed.
 
%  However, before  going into  the details a  cautionary remark  is in
%   order.  In  particular, we  emphasize that  small variations  in the
%   values  of  the parameters  adopted  for  the different  simulations
%   presented below can lead to substantially different outcomes.

%%%%%%%%%%%%%%%%%%%%%%%%%%%%%%%%%%%%%%%%%%%%%%%%%
\subsection{Accretion rate set by the cooling timescale}
\label{sec:tcool}
%%%%%%%%%%%%%%%%%%%%%%%%%%%%%%%%%%%%%%%%%%%%%%%%%

\begin{figure}
\centering
\includegraphics[width=\hsize,clip]{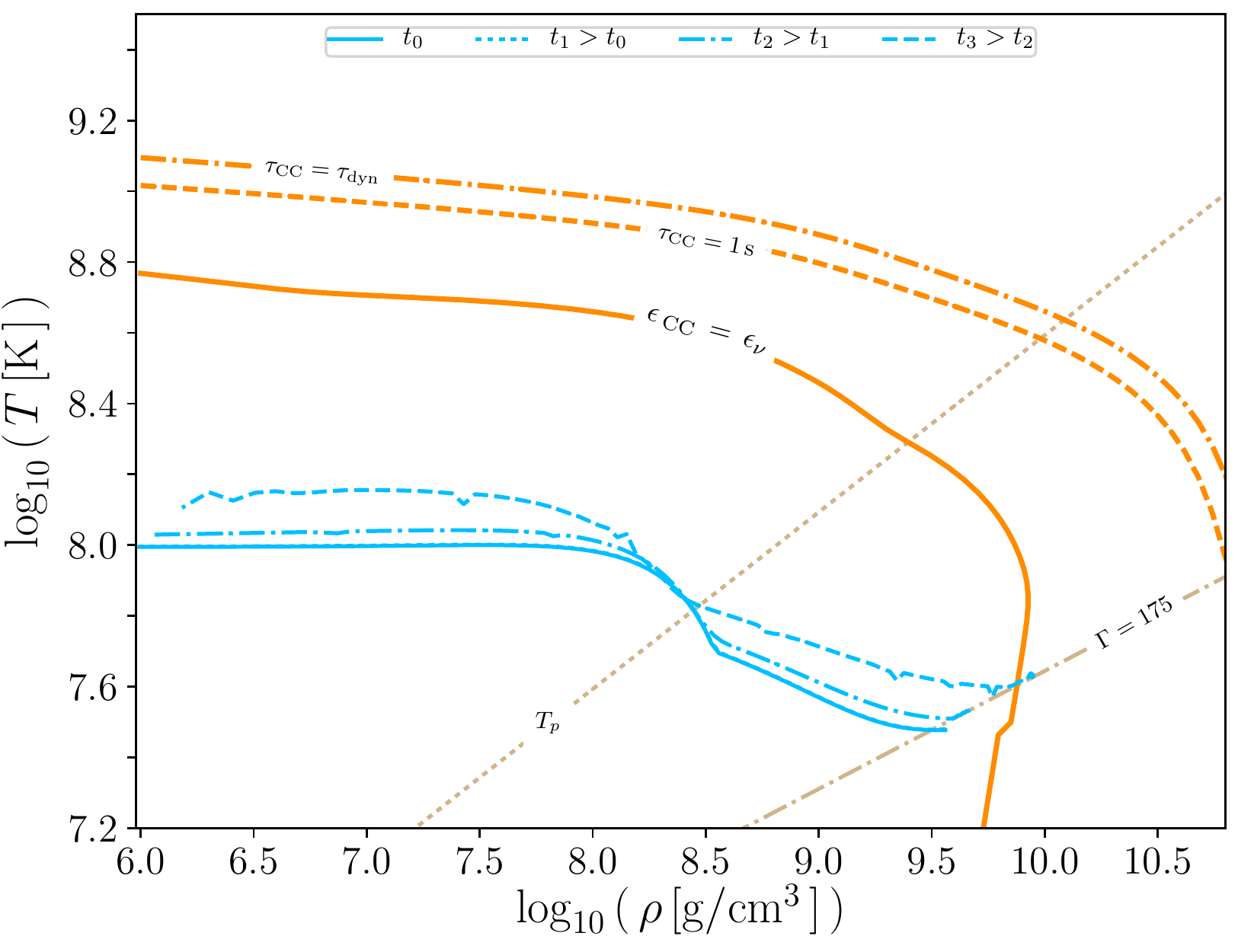}
\caption{Temperature-density profiles  of the  central white  dwarf at
  different  times of  the  evolution  for model  A.   This model  was
  computed  assuming  $B=10^9$~G and  an  accretion  rate set  by  the
  cooling timescale.  We  also show the carbon  ignition line, labeled
  as   $\epsilon_{\rm  CC}=\epsilon_\nu$,   and  two   carbon  burning
  timescales  $\tau_{\rm CC}=\tau_{\rm  dyn}$  and $\tau_{\rm  CC}=1\,
  {\rm s}$.  In  these panels the crystallization curve  is labeled as
  $\Gamma=175$, and the plasma temperature as $T_{\rm p}$. The configuration at $t_1$ has a temperature profile almost equal to the initial one. At $t_2$ and $t_3$, it heats at the center and at the surface. }
\label{fig:T_EddLimit}
\end{figure}

To start with we discuss our  reference model. This is model A
  in Table~\ref{tab:InConModels}. For this model we adopted a magnetic
  field  $B=10^9$~G. We  note  that  during the  early  phases of  the
  evolution of the post-merger  remnant the dominant cooling mechanism
  of  the external  layers  of  the central  white  dwarf is  neutrino
  emission.  Therefore, the initial  relevant timescale to compute the
  accretion   rate   is  $\dot{M}_{\rm   WD}=M_{\rm   WD}/\tau_\nu\sim
  10^{20}$~g~s$^{-1}$.  This  accretion rate  is of  the order  of the
  Eddington limit. However,  in our calculations the  evolution of the
  model is followed  self-consistently. That is, we  computed the mass
  accretion rate with the cooling  timescale provided by the evolution
  -- see below.

\begin{figure}
\centering
\includegraphics[width=\hsize,clip]{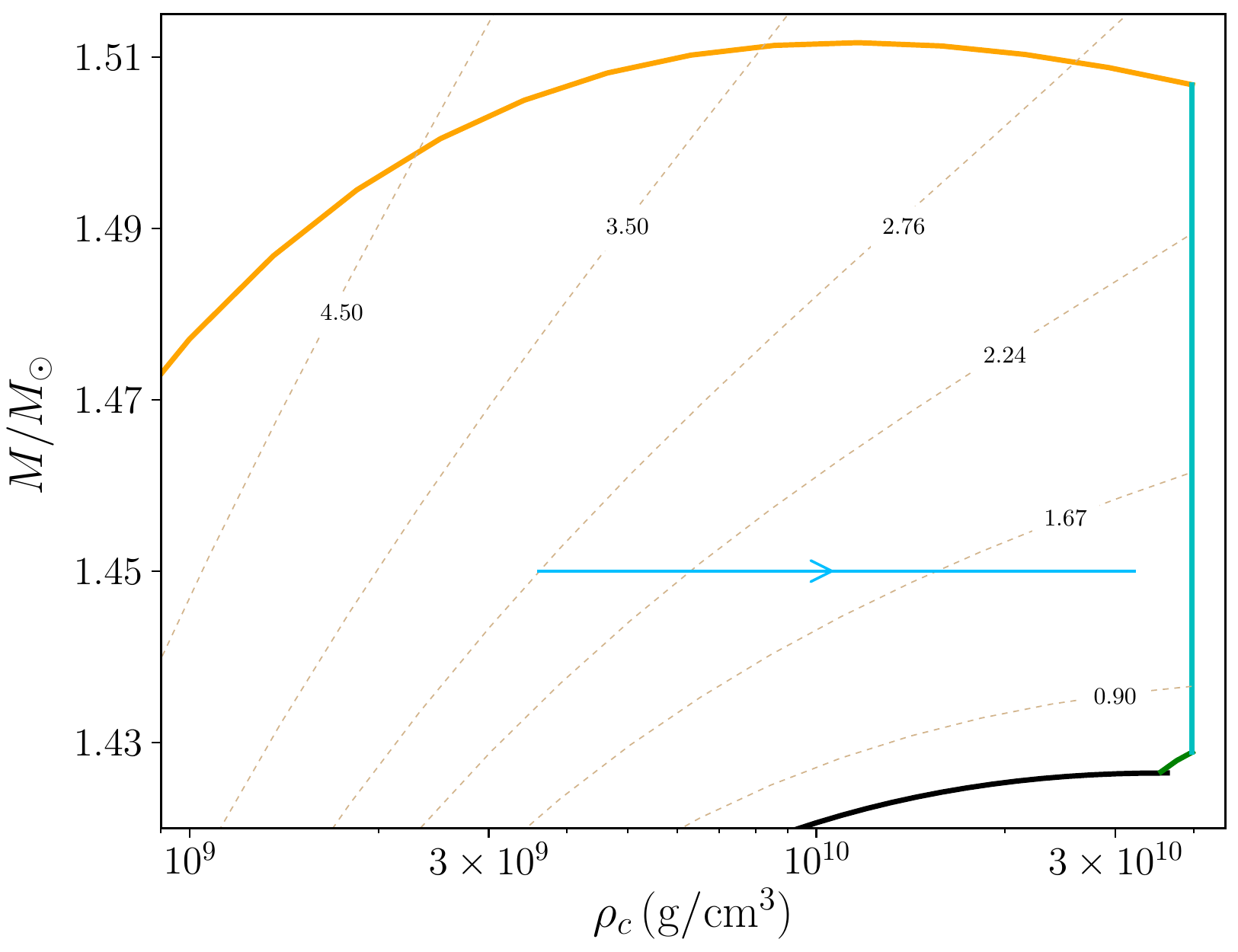}
\caption{Evolution  of  model A  in  the  mass-central density  plane.
  model A at  selected times, see text for details.  The colors of the
  lines indicating the regions  of the various instabilities discussed
  here are the same shown in Figure~\ref{fig:WDrot}.  Also shown are the
  contours of constant angular momentum (dotted lines).}
\label{fig:modelAtrack}
\end{figure}

For  this   model,  the   magnetospheric  radius   is  $R_{\rm
    mag}=0.062\,  R_{\sun}$  and  the  corotation  radius  is  $R_{\rm
    co}=0.0039\,  R_{\sun}$.   Thus,  $R_{\rm  mag}>R_{\rm  co}>R_{\rm
    WD}$,  so the  dipole radiation  torque and  the propeller  torque
  drive  the  evolution of  the  spin  of  the remnant.   Under  these
  conditions  the   central  object  does  not   accrete  matter.   To
  illustrate  the  evolution,   Figure~\ref{fig:T_EddLimit}  shows  some
  temperature profiles at selected times. In particular, we show these
  profiles at  times $t_0$  -- corresponding to  the beginning  of the
  evolution, solid  line -- at  a time  just before the  central white
  dwarf reaches explosive conditions -- time $t=t_3$, long dashed line
  -- and at  two intermediate stages  -- times $t_1$ and  $t_2$, short
  dashed line and dashed-dotted line,  respectively. As can be seen in
  this figure,  during the entire  evolution both the  central regions
  and the outer  layers of the white dwarf are  compressed and heated.
  These two  regions of the white  dwarf are separated by  the line at
  which  the temperature  of  the nearly  isothermal  core equals  the
  plasma temperature, $T_{\rm P}$.  This  behavior is a consequence of
  the torques acting on the white  dwarf. The acting torques brake the
  white dwarf, decreasing its angular velocity.  As a consequence, the
  centrifugal force  decreases and the gravitational  force dominates.
  Thus, to balance the enhanced gravity the density increases, so does
  the temperature.   Ultimately, the  center of  the star  reaches the
  thermodynamic conditions  needed to burn carbon  explosively.  These
  conditions   are  illustrated   in   this  figure   by  the   curves
  $\epsilon_{\rm  nuc}  =  \epsilon_{\nu}$,  $\tau_{\rm  CC}=\tau_{\rm
    dyn}$ and $\tau_{\rm CC}=1$~s. This occurs at time $40.8$~yr, just
  before  these regions  reach the  beta-decay instability  limit, and
  when    the   central    regions   of    the   star    are   already
  crystallizing. Figure~\ref{fig:modelAtrack} displays  the evolution of
  model A  in the  mass-central density  plane.  As  can be  seen, the
  evolutionary track corresponds to pure compression, and no matter is
  accreted. Note  as well that  carbon in  the central regions  of the
  star  is ignited  before  the inverse  $\beta$-decay instability  is
  reached.

\begin{figure}
\centering
\includegraphics[width=\hsize,clip]{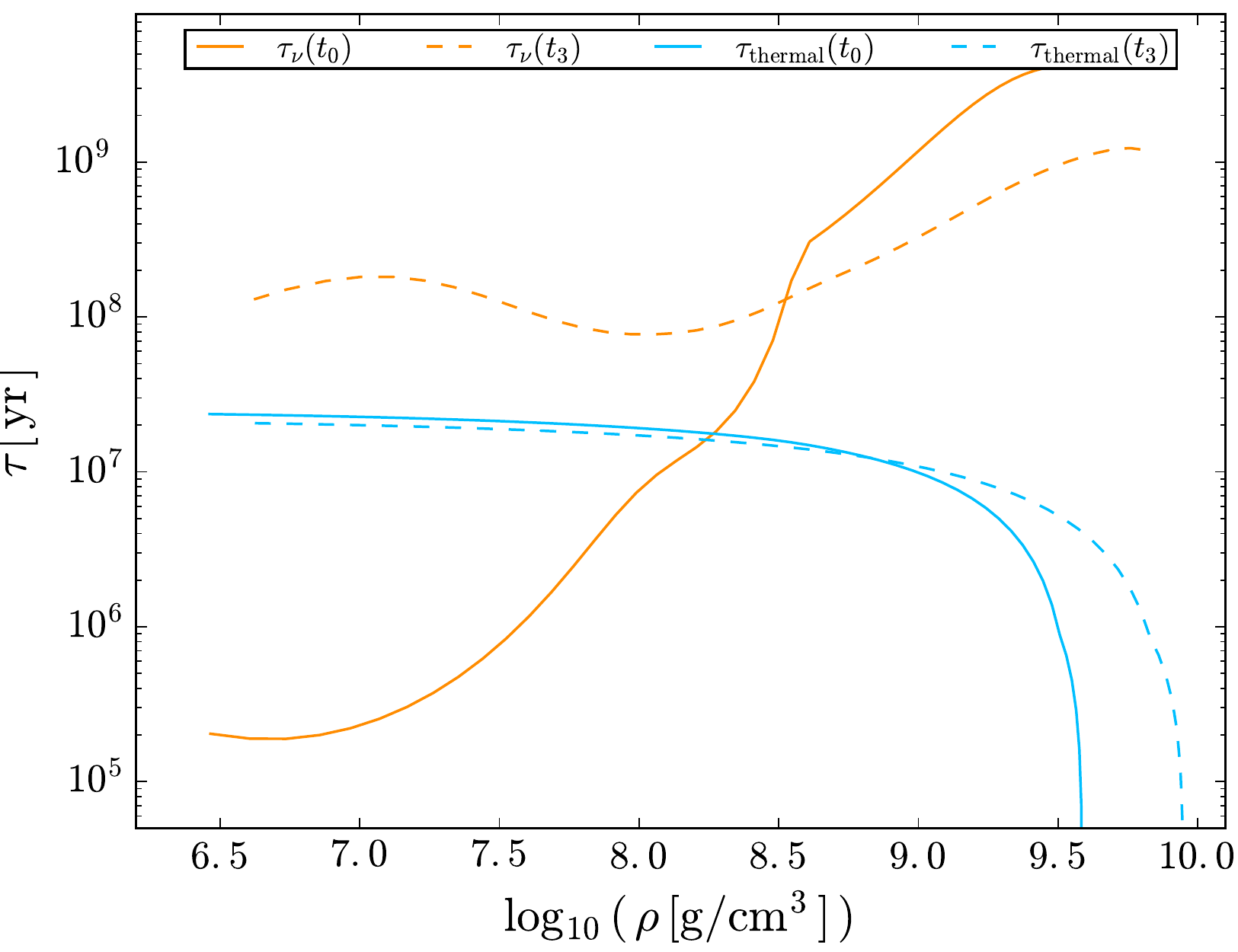}
\caption{Run of the neutrino and thermal timescales in the interior of
  model A at selected times, see text for details.}
\label{fig:Ttaus_EddLimit}
\end{figure}

  Figure~\ref{fig:Ttaus_EddLimit}  shows the  run of  the neutrino
  and thermal timescales in the white dwarf interior of model A at two
  relevant times, namely  at times $t=t_0$ and  $t=t_3$, as previously
  defined.  As  can be seen,  for $t=t_0$ neutrinos cool the external
  layers of the star, while  in the internal regions thermal diffusion
  is the  dominant transport  mechanism.  At  time $t_3$  the relevant
  cooling  timescale in  the outer  layers of  the white  dwarf is  no
  longer the neutrino timescale, but the thermal timescale.  Actually,
  it is important to realize that in the dense inner core of the white
  dwarf the shorter timescale is  always the thermal timescale.  Thus,
  Figure~\ref{fig:Ttaus_EddLimit}   clearly  shows   that  the   cooling
  timescale must be computed  self-consistently along the evolution of
  the remnant to  obtain physically sound results.   We emphasize that
  although this  plot illustrates  these timescales  for model  A, our
  calculations  demonstrate that  this is  a representative  case, and
  that   for  all   the  models   listed   in  the   top  section   of
  Table~\ref{tab:InConModels} the run of these timescales is similar.

\begin{table*}
\centering
\caption{Carbon ignition location and  evolutionary times  of the  post-merger remnants
  for  several  values of  the  magnetic  field strength  and  initial
  angular velocity.}
\begin{tabular}{cccccccccl}
  \hline
  \hline
  Model &  $B_{\rm WD}$~[G] & \multicolumn{7}{c}{$\Delta\, t$ [yr] } & Carbon ignition \\
  \cline{3-9}
    $\Omega_{\rm WD}\; [{\rm s}^{-1}]$ & & 2.3& 2.5 & 2.6 & 2.7 & 2.8 & 2.9 & 3.0\ & \\ 
 \hline
 A & $10^9$ & $5.50\times 10^1$ & $4.68\times 10^1$ & $4.33\times 10^1$ & $4.02\times 10^1$ & $4.71\times 10^1$ & $6.95\times 10^1$ & $4.08\times 10^1$ & Center\\
\hline
 B & $10^6$ & $4.00\times 10^3$ & $8.81\times 10^3$ & $6.39\times 10^3$ & $5.27\times 10^3$ & $5.63\times 10^3$ & $3.65\times 10^3$ & $4.94\times 10^3$ & Off-center\\
 C & $10^7$ & $2.81\times 10^4$ & $3.95\times 10^4$ & $2.09\times 10^4$ & $1.66\times 10^4$ & $1.34\times 10^4$ & $1.59\times 10^4$ & $1.50\times 10^4$ & Center\\
 D & $10^8$ & $9.29\times 10^2$ & $7.96\times 10^2$ & $7.39\times 10^2$ & $8.92\times 10^2$ & $8.31\times 10^2$ & $7.76\times 10^2$ & $5.61\times 10^2$ & Center\\
 \hline
 \hline
\end{tabular}
\label{tab:Times}
\end{table*}

\begin{figure}
\centering
\includegraphics[width=\hsize,clip]{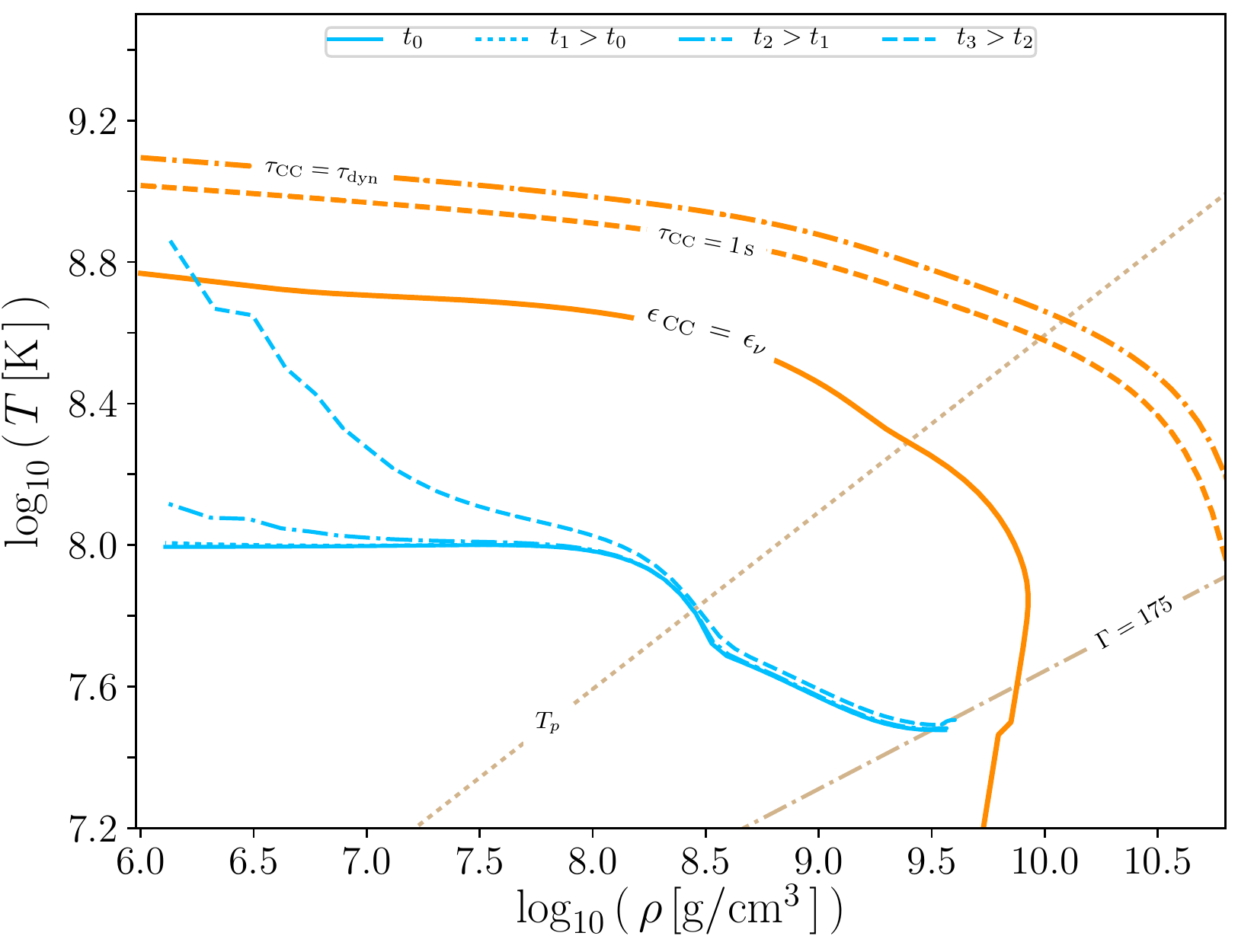}
\caption{Same as Figure~\ref{fig:T_EddLimit} but for the case in which a
  modest magnetic  field, $B=10^6$~G, is  adopted. This is model  B in
  Table~\ref{tab:InConModels}.}
\label{fig:T_EddLimit2}
\end{figure}

 To  study  the  dependence  of the  evolution  of  the  merged
  configuration on  the magnetic field we  ran a suite of  models with
  varying strengths of the magnetic field.   These are models B, C and
  D  in Table~\ref{tab:InConModels},  for  which  we adopted  magnetic
  field strengths $B=10^6$, $10^7$ and $10^8$~G, respectively, keeping
  unchanged the  rest of the parameters  of model A.  For  the sake of
  conciseness  we  only discuss  model  B,  which corresponds  to  the
  smallest magnetic field strength, $10^6$~G.

\begin{figure}
\centering
\includegraphics[width=\hsize,clip]{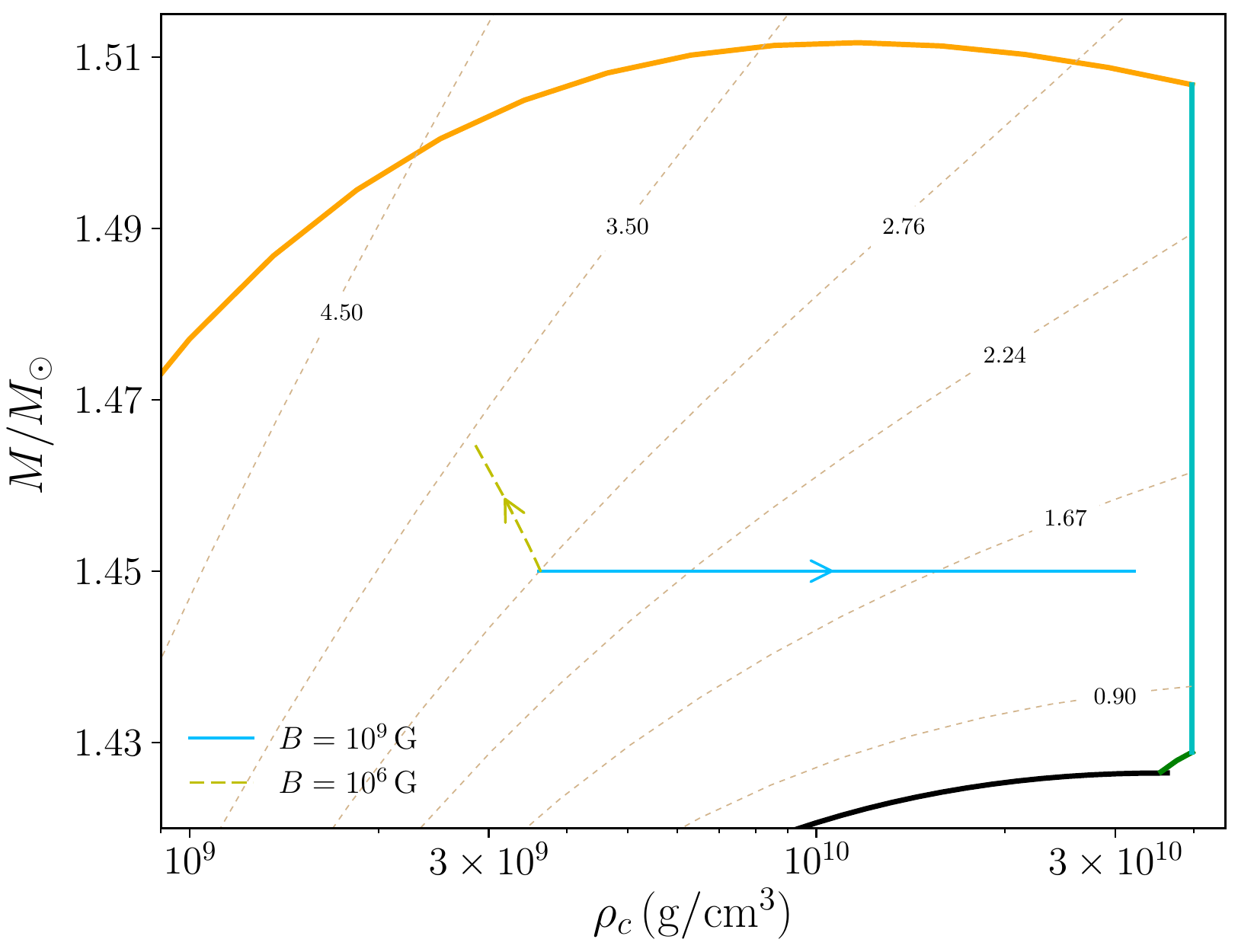}
\caption{Same as Figure~\ref{fig:modelAtrack} but including the case in which a magnetic  field, $B=10^6$~G, is  adopted (yellow line). This is model B in   Table~\ref{tab:InConModels}.}
\label{fig:modelBtrack}
\end{figure}

 Figure~\ref{fig:T_EddLimit2} shows the  evolution of the remnant
  at different times for this  model.  These times were selected using
  the same  criteria we used  previously, and  the line coding  is the
  same. In  this case $R_{\rm  mag}<R_{\rm WD}$ and the  central white
  dwarf can accrete  matter from the disk.  Under  these conditions we
  find that accretion  onto the white dwarf heats the  outer layers of
  the  star,  while  the  temperature   of  its  core  remains  almost
  unchanged. Ultimately the  very outer layers of the  white dwarf are
  heated to such an extent that carbon is ignited off-center. However,
  an  inspection   of  Figure~\ref{fig:modelBtrack},  which   shows  the
  evolution in  the mass-central density plane,  reveals that actually
  the central regions  of the star expand slightly. This  again is due
  to the acting torques, that spin-up the white dwarf.

  We found  that if  the magnetic  field of  the white  dwarf is
  higher than $3.84\times10^6$~G, the  magnetospheric radius is larger
  than the co-rotation radius.  Thus, in this case the central remnant
  does not  accrete more material  after the merger, so  the propeller
  torque and  the dipole radiation  torque drive the evolution  of its
  spin rate.  On  the other hand, if the magnetic  field is lower than
  $1.88\times 10^6$~G,  the magnetospheric radius is  smaller than the
  white  dwarf  radius.   Hence,  the  central  white  dwarf  accretes
  material.  Consequently, in this case  the evolution of the rotation
  rate is  driven by  the accretion torque.   Thus, for  these initial
  conditions the magnetic field determines  the accretion rate and the
  evolution of the star.

 To further study the dependence  of these results on the input
  parameters, we  also analyzed the  dependence on the  initial angular
  velocity.  In Table~\ref{tab:Times}  we list the times  at which the
  post-merger  central  star reaches  the  conditions  suitable for  a
  explosion to occur,  for several magnetic field  strengths (models A
  to D) and initial angular velocities.  Clearly, the duration of this
  phase as  well as the final  outcome depend sensitively on  value of
  the adopted  magnetic field, and to  a lesser extent on  the initial
  angular velocity.

%%%%%%%%%%%%%%%%%%%%%%%%%%%%%%%%%%%%%%%%%%%%%%%%%
\subsection{Accretion rate set by viscous timescale}
\label{sec:tvisc}
%%%%%%%%%%%%%%%%%%%%%%%%%%%%%%%%%%%%%%%%%%%%%%%%%

In the  previous section  we have  assumed that  the accretion
  rate on  the white dwarf  is given by  the shorter timescale  of the
  cooling mechanisms. In  this section we present the  results for the
  case in which the white dwarf accretes material at a rate set by the
  viscous timescale of the Keplerian disk.

\begin{figure}[t]
\centering
\includegraphics[width=0.9\hsize,clip]{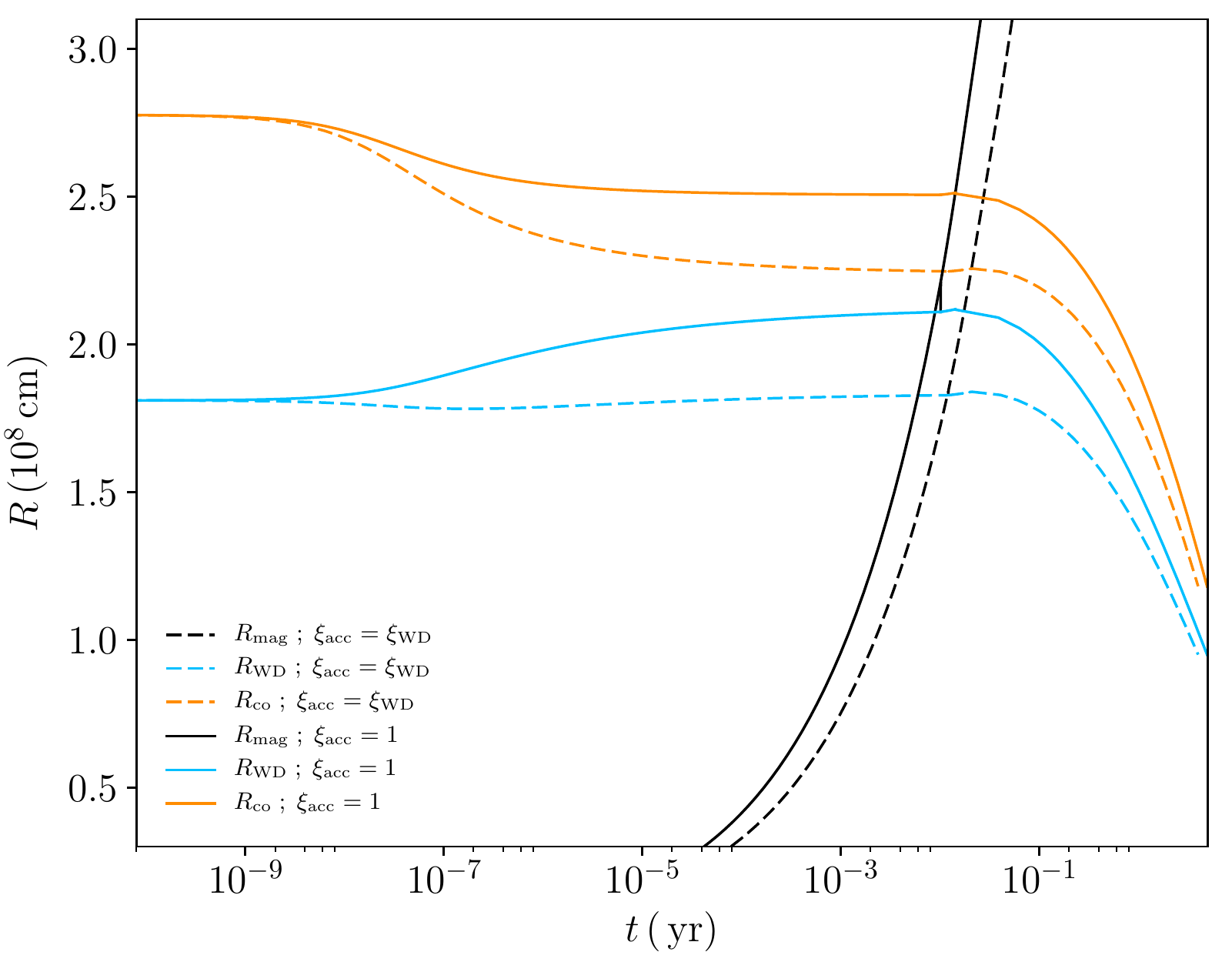}\\
\includegraphics[width=0.9\hsize,clip]{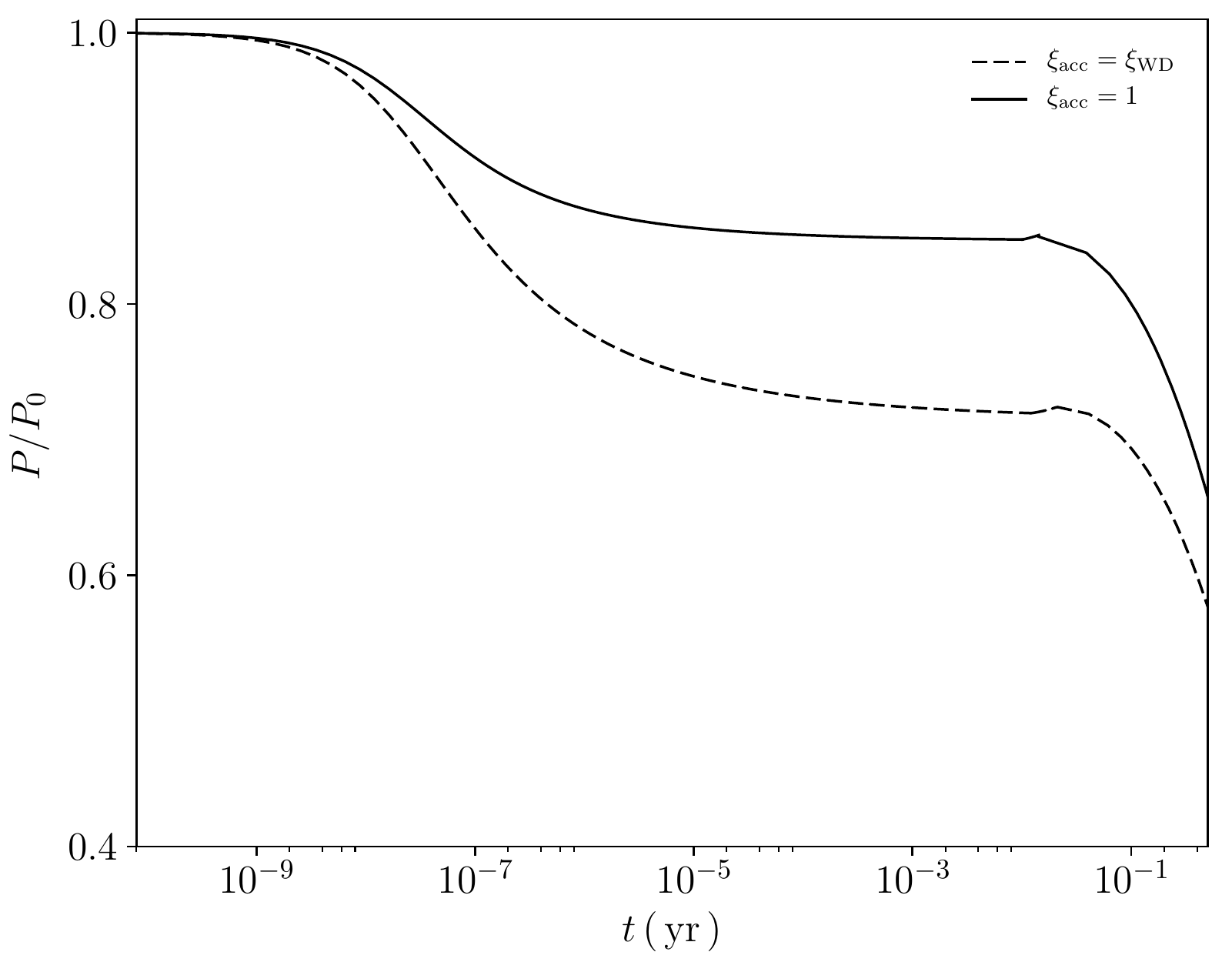}
\caption{The  top panel  shows  the  evolution of  the  radius of  the
  remnant, $R_{\rm  WD}$, the corotation  radius $R_{\rm co}$,  and the
  magnetospheric radius, $R_{\rm  mag}$ for model E,  while the bottom
  panel  shows the  evolution of  the rotation  period of  the central
  white dwarf.}
\label{fig:WDevol_1}
\end{figure}

\begin{figure}
\centering
\includegraphics[width=0.9\hsize,clip]{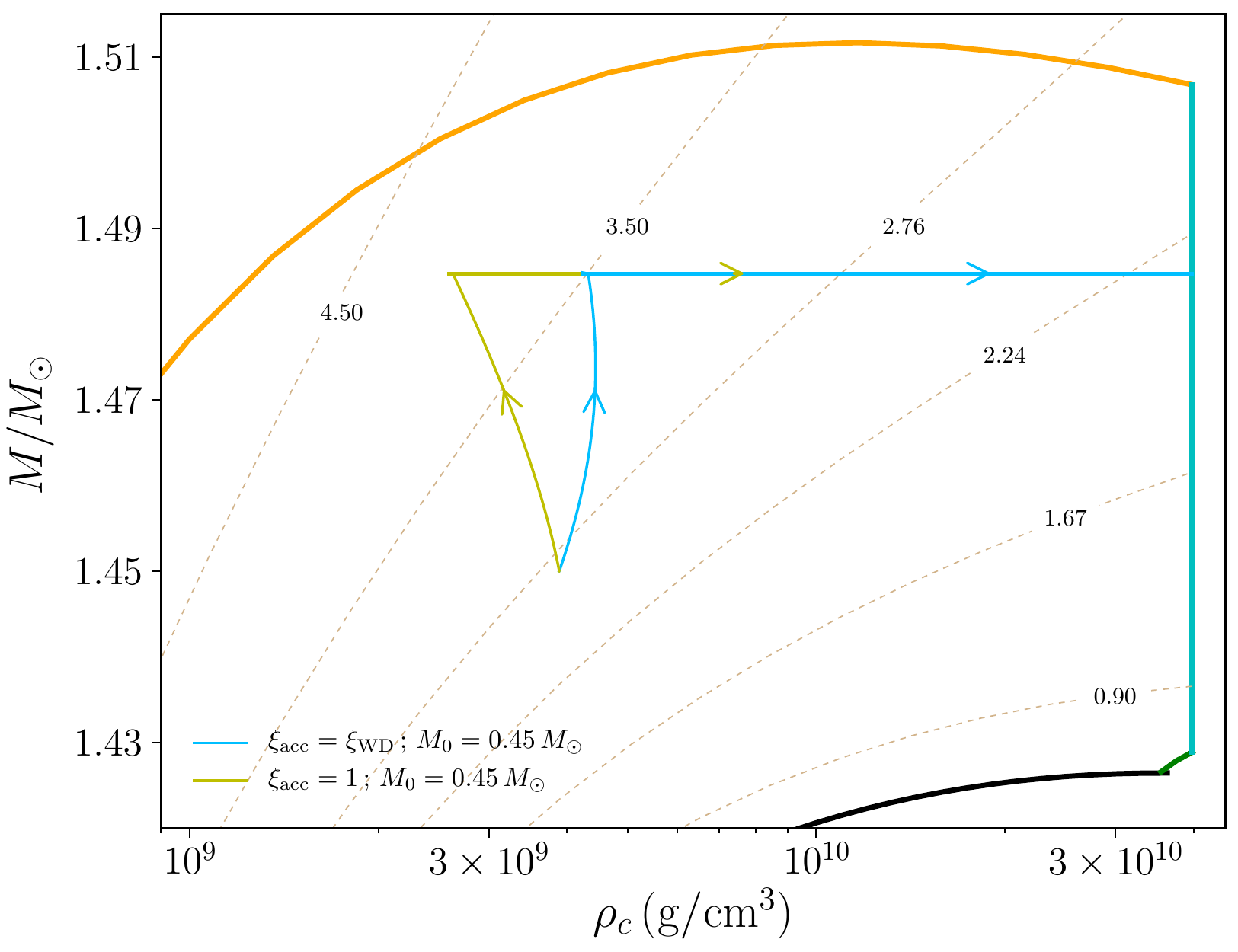}
\caption{Evolutionary tracks in the mass-density plane for model E for
  two   assumptions  about   the   value  of   $\xi_{\rm  acc}$,   and
  $\varepsilon=0.1$.}
\label{fig:WDevol_2}
\end{figure}

For the  sake of definiteness in  the following we discuss  in detail,
with the help of  Figures~\ref{fig:WDevol_1} and \ref{fig:WDevol_2}, the
time evolution of model E, which is similar to model A, the
only difference being that in this case the accretion rate is computed
adopting the viscous  timescale, while the rest  of initial conditions
and assumptions are the exactly the same.  Furthermore, for this model
we  study two  possibilities  for the  efficiency parameter  $\xi_{\rm
acc}$  in  Equation~(\ref{eq:Tacc}).   The  first  of  these  possibilities
corresponds to  the case in  which $\xi_{\rm acc}=1.0$, that  is, when
matter is accreted  on the remnant with the Keplerian  velocity at the
inner disk  radius.  The second one  corresponds to the case  in which
matter is accreted  on the remnant with the Keplerian  velocity at the
surface of the remnant, which we label as $\xi_{\rm acc}=\xi_{\rm WD}$
in the figures discussed below.

Shortly after  the merger,  due to the  high accretion  rates, $R_{\rm
mag}< R_{\rm  WD}$ --  see the  top panel  of Figure~\ref{fig:WDevol_1}.
Hence, the  only torque acting  on the  white dwarf is  that resulting
from accretion.   For the set  of parameters  of model E,  the initial
accretion rate onto  the white dwarf given  by Equation~(\ref{eq:Mdisk}) is
large, even  if  an efficiency  of $\varepsilon=0.1$  is  adopted,
$\dot{M}_{\rm WD}\sim 0.01\,M_\sun$~s$^{-1}$.   Consequently, the mass
of the white  dwarf rapidly increases during this  phase. However, its
central density only increases in the case in which matter is accreted
with  the  Keplerian   velocity  of  the  remnant  --   blue  line  in
Figure~\ref{fig:WDevol_2}  -- whereas  in  the case  in which  $\xi_{\rm
acc}=1.0$ -- green  line in this figure -- the  central density of the
white dwarf decreases.

The evolution of the central density of the remnant  is the result of
an intricate trade-off between the increase in mass, the change in the
rotation period  due to the  acting torque,  and the evolution  of the
moment of inertia of the remnant.  The increase in mass of the remnant
alone clearly would result in an  increase of the central density. The
change in the  rotation speed due to the acting  torque -- which tends
to increase the angular velocity --  would result in a decrease of the
central density,  as the centrifugal force  increases.  Finally, since
$I\propto M_{\rm  WD} R_{\rm WD}^2$,  as the  mass of the  white dwarf
increases  due to  accretion, the  moment of  inertia would  increase.
However, since the mass-radius  relationship for rotating white dwarfs
depends crucially on the angular  velocity, the radius of the remnant,
hence $I$, ultimately depends on  the acting torque.  Furthermore, for
super-Chandrasekhar  white   dwarfs  the  slope  of   the  mass-radius
relationship is very steep. Consequently small changes in the mass can
induce large variations  of the radius of the  remnant.  The interplay
between these factors is rather complex,  but in general terms we find
that  the evolution  of  the moment  of inertia  is  dominated by  the
variation of the radius of the remnant.

\begin{figure}
\centering
\includegraphics[width=0.9\hsize,clip]{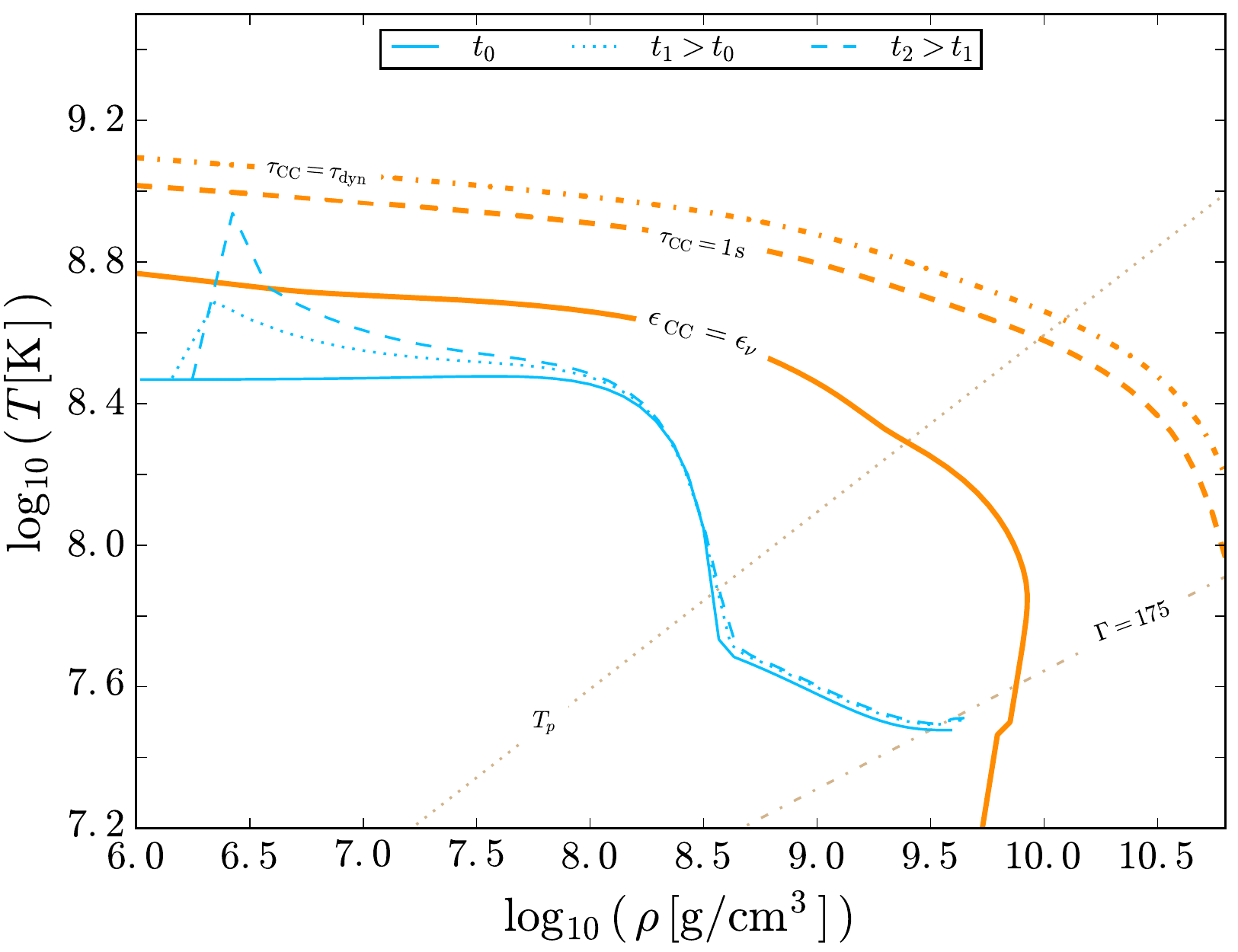}\\
\includegraphics[width=0.9\hsize,clip]{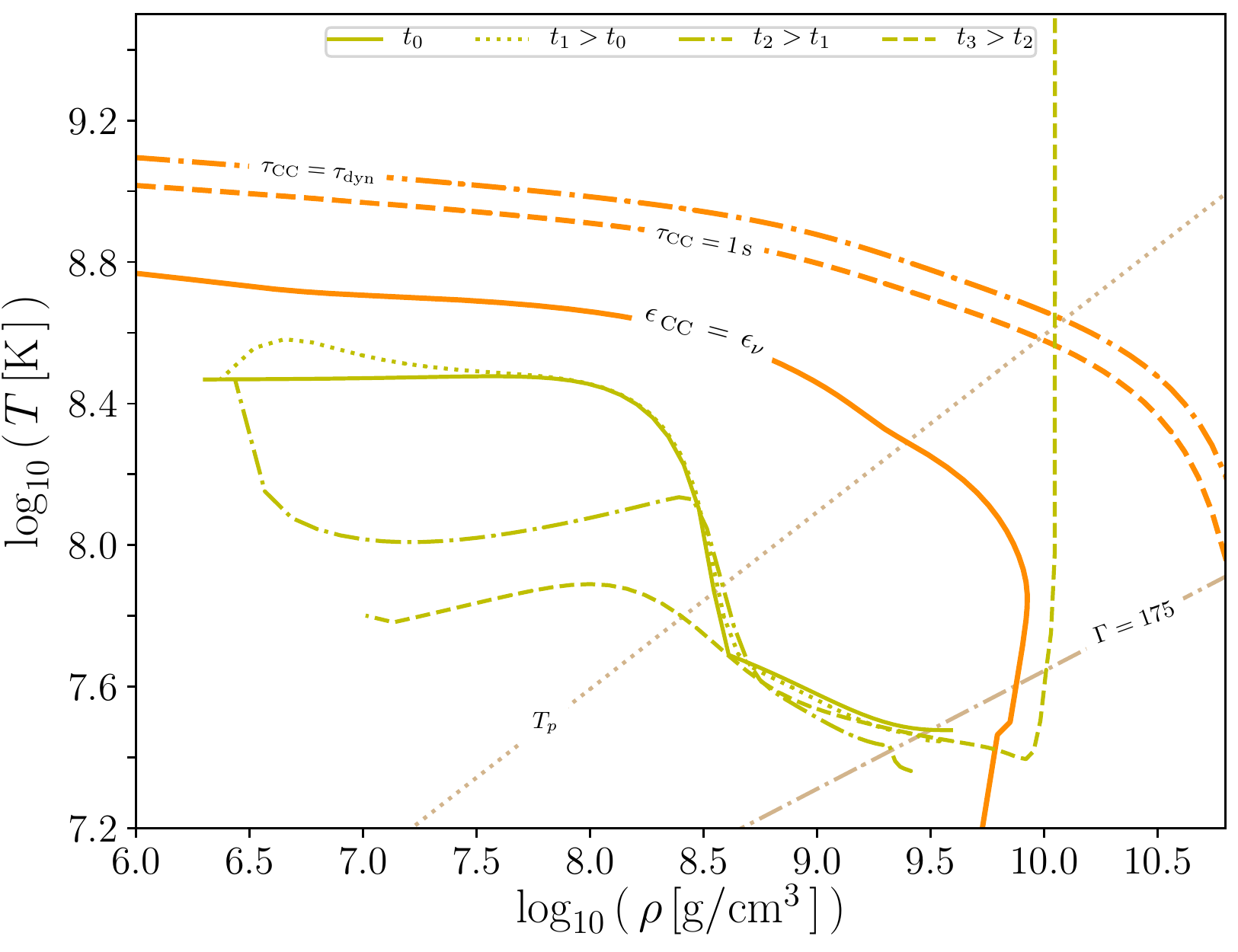}
\caption{Temperature-density profiles  of the  central white  dwarf at
  different times  of the evolution for  model E. The top  panel shows
  the model with $\xi=\xi_{\rm acc}$,  while the bottom panel displays
  the model for which $\xi=1$ is  adopted. Time $t_1$ corresponds to a
  time  shortly after  accretion from  the disk  starts (short  dashed
  line), $t_2$ is the time at which the merged remnant enters into the
  propeller phase  and accretion stops  (dashed line), and $t_3$  is a
  time just before the post-merger  object crosses the beta-instability limit %carbon ignition line
  (long dashed line).}
\label{fig:WDevol_Trhoprofile}
\end{figure}

It  is  then  clear  that  the crucial  parameter  that  dictates  the
evolution of  the central density  of the remnant is  $\xi_{\rm acc}$.
Specifically,  when  $\xi_{\rm acc}=1.0$  is  adopted  the remnant  is
spun-up very  rapidly by the  acting torque -- as  it is shown  in the
bottom  panel  of  Figure~\ref{fig:WDevol_1}  --  its  radius  increases
notably -- see the  top panel of this figure --  the moment of inertia
increases markedly, and  the central density first  decreases.  On the
contrary, if  matter is accreted onto  the surface of the  white dwarf
with the  same angular velocity of  the remnant, the acting  torque is
considerably  smaller, the  moment  of  inertia remains  approximately
constant, the radius of the remnant decreases, and the central density
increases steadily.  In  both cases, once the  magnetic radius becomes
larger  than  the  radius  of  the  white  dwarf  ($R_{\rm  WD}<R_{\rm
mag}<R_{\rm co}$),  the dipole and the  disk-interaction torques drive
the   evolution    of   the    remnant   --    the   top    panel   of
Figure~\ref{fig:WDevol_1} clearly depicts this.   When this happens, the
fastness parameter  of the white  dwarf is $\sim  0.8$.  Consequently,
shortly after ($\la  0.5\, {\rm yr}$), the remnant  reaches a fastness
parameter  $\omega_{\rm f}=1$,  and  the evolution  is  driven by  the
propeller torque.  At this point  of the evolution, accretion from the
disk  stops.   Consequently, the  moment  of  inertia of  the  remnant
decreases considerably, the rotation period decreases as well, and the
white dwarf  contracts.  Hence, the central  density increases rapidly 
(see the top panel of  Figure~\ref{fig:WDevol_1}).  All this sequence of
events  ultimately leads  the remnant  to  cross the  line of  inverse
$\beta$-decay instability.  It is interesting to note that the mass of
the remnant  when it crosses the  instability line does not  depend on
the adopted valued of $\xi_{\rm acc}$. This depends on the mass-accretion rate onto the white draft, and consequently on the adopted efficiency parameter $\xi$. The parameter $\xi_{\rm acc}$ is directly related to the change of angular momentum. Thus $\xi_{\rm acc}$ and $\xi$ play different roles in the evolution of the post-merger configuration, although its effect is combined as shown by Equation~(\ref{eq:Tacc}). 

We now address our attention to thermal evolution  of the post-merger
  remnant  of   model  E.    Figure~\ref{fig:WDevol_Trhoprofile}  shows
several  temperature-density  profiles  at selected  times  after  the
merger took  place.  The top  panel of this  figure shows the  case in
which  $\xi=\xi_{\rm  acc}$  is  adopted,  whereas  the  bottom  panel
corresponds to  the case  in which $\xi=1$  is employed.   The initial
temperature profiles  are shown  as a  solid blue  and green  lines in
Figure~\ref{fig:WDevol_Trhoprofile}, respectively. In both cases, during
the  first evolutionary  phases  the large  accretion rates  discussed
earlier heat the outer layers of the star.  However, neutrino emission
also plays  a significant role.   When $\xi=\xi_{\rm acc}$  is adopted
compressional work  exceeds neutrino emission, and  thus an off-center
temperature  peak  rapidly  grows.   In  particular,  the  temperature
profile peaks  at $\log\rho\sim  6.54$.  As time  passes by,  the peak
temperature increases,  and ultimately the  outer layers of  the white
dwarf  reach  the thermodynamic  conditions  needed  to ignite  carbon
explosively.

On the  contrary, when $\xi=1.0$  is adopted (bottom  panel of
  Figure~\ref{fig:WDevol_Trhoprofile}) the evolution is more complex. In
  this  case accretion  first heats  the outer  layers of  the central
  white dwarf,  but neutrino cooling  dominates. Hence, after  a short
  time interval the entire central  white dwarf cools to a temperature
  smaller  than the  initial  one. Time  $t_2$ is  the  time at  which
  accretion stops. Thus, for times  longer than $t_2$ the evolution of
  the star  is driven by  angular momentum looses by  dipole emission.
  Eventually, at  time $t_3$ carbon  is ignited  at the center  of the
  white dwarf.

%%%%%%%%%%%%%%%%%%%%%%%%%%%%%%%%%%%%%%%%%%%%%%%%%
\subsubsection{Sensitivity of the results to the free parameters}
\label{sec:sensitivity}
%%%%%%%%%%%%%%%%%%%%%%%%%%%%%%%%%%%%%%%%%%%%%%%%%

\begin{figure}
\centering
\includegraphics[width=0.9\hsize,clip]{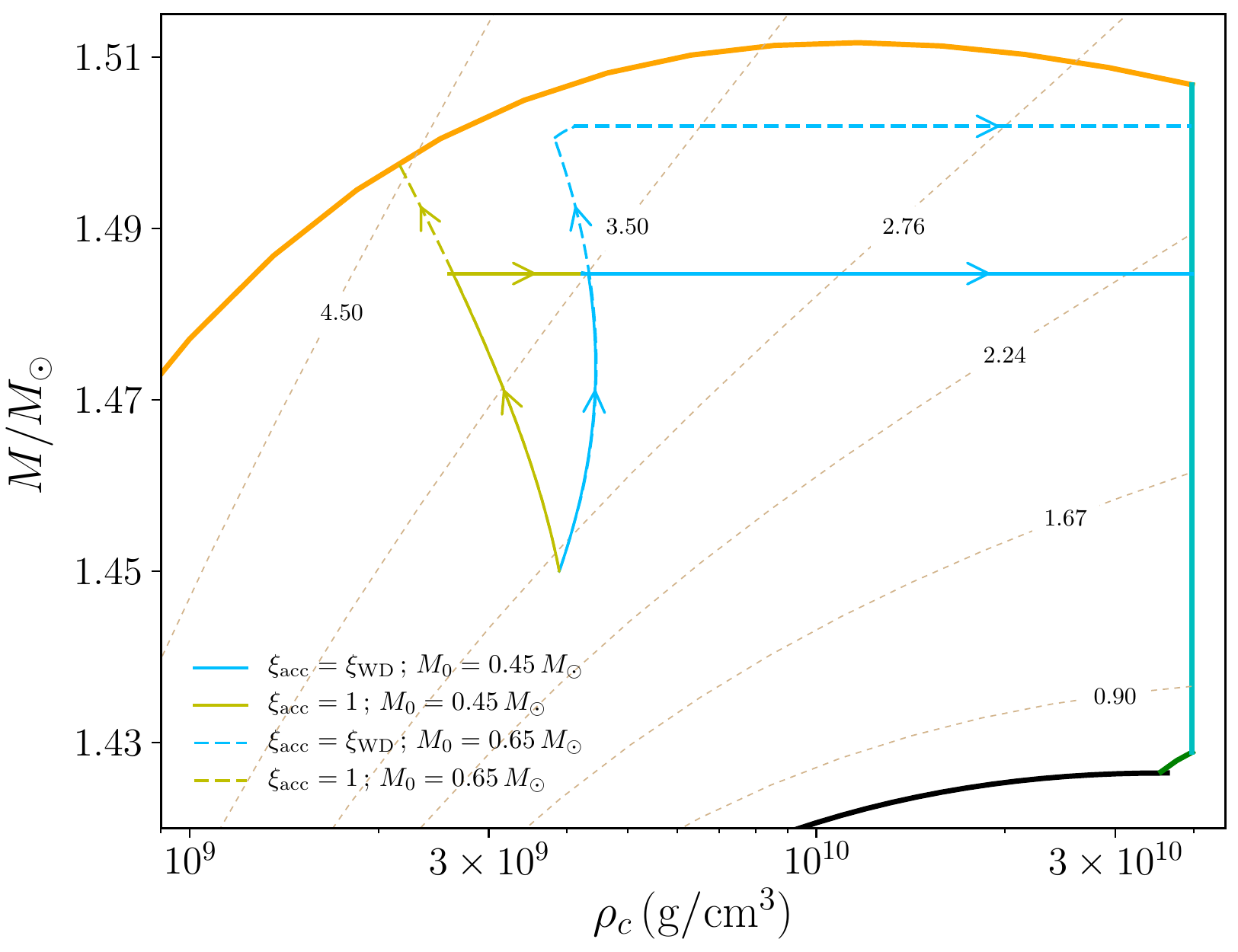}
\caption{Same as  Figure~\ref{fig:WDevol_2}. We compare  models E  (solid lines)
  and F (dashed lines) for two different disk initial mass, $0.45\,M_\odot$ and $0.65\,M_\odot$, respectively.}
\label{fig:WDevol_diskmass}
\end{figure}

We first compare  the evolution when different masses of  the disk are
adopted, keeping the mass of the remnant fixed.  With this in mind, in
Figure~\ref{fig:WDevol_diskmass}  we  show  the evolutionary  tracks  of
model  F  and  compare  them  with  those  of  model  E.   This  model
corresponds to a final post-merger  remnant of mass $1.45 \, M_{\sun}$
and an  initial disk mass $0.65\,  M_{\sun}$, whereas the rest  of the
parameters    of    the    model    were    not    varied    --    see
Table~\ref{tab:InConModels}.  Additionally,  as we did for  model E we
also studied two possibilities.  The  first of these corresponds to an
accretion efficiency  parameter $\xi_{\rm  acc}=1.0$, whereas  for the
second  one we  assumed that  at the  inner disk  radius the  accreted
material has  the angular velocity  of the white  dwarf, $\xi_{\rm acc}=\xi_{\rm
  WD}$. Note that, in the case in which $\xi_{\rm acc}=0.1$ is adopted,  the evolution of this model is very similar to that
of model E, but it arrives to the mass-shedding limit in the accretion phase. When
$\xi_{\rm acc}=\xi_{\rm WD}$  is chosen,  the remnant  also crosses  the line  of inverse
$\beta$-decay instability.  The  only difference is that  for model F,
which  has a  disk considerably  more massive,  more mass  is accreted
before the remnant  crosses the instability line.   Hence, the remnant
has  a  larger   mass  in  the  propeller  phase,  and   the  line  of
$\beta$-decay  instability is  crossed when  the white  dwarf is  more
massive.

\begin{figure}
\centering
\includegraphics[width=0.9\hsize,clip]{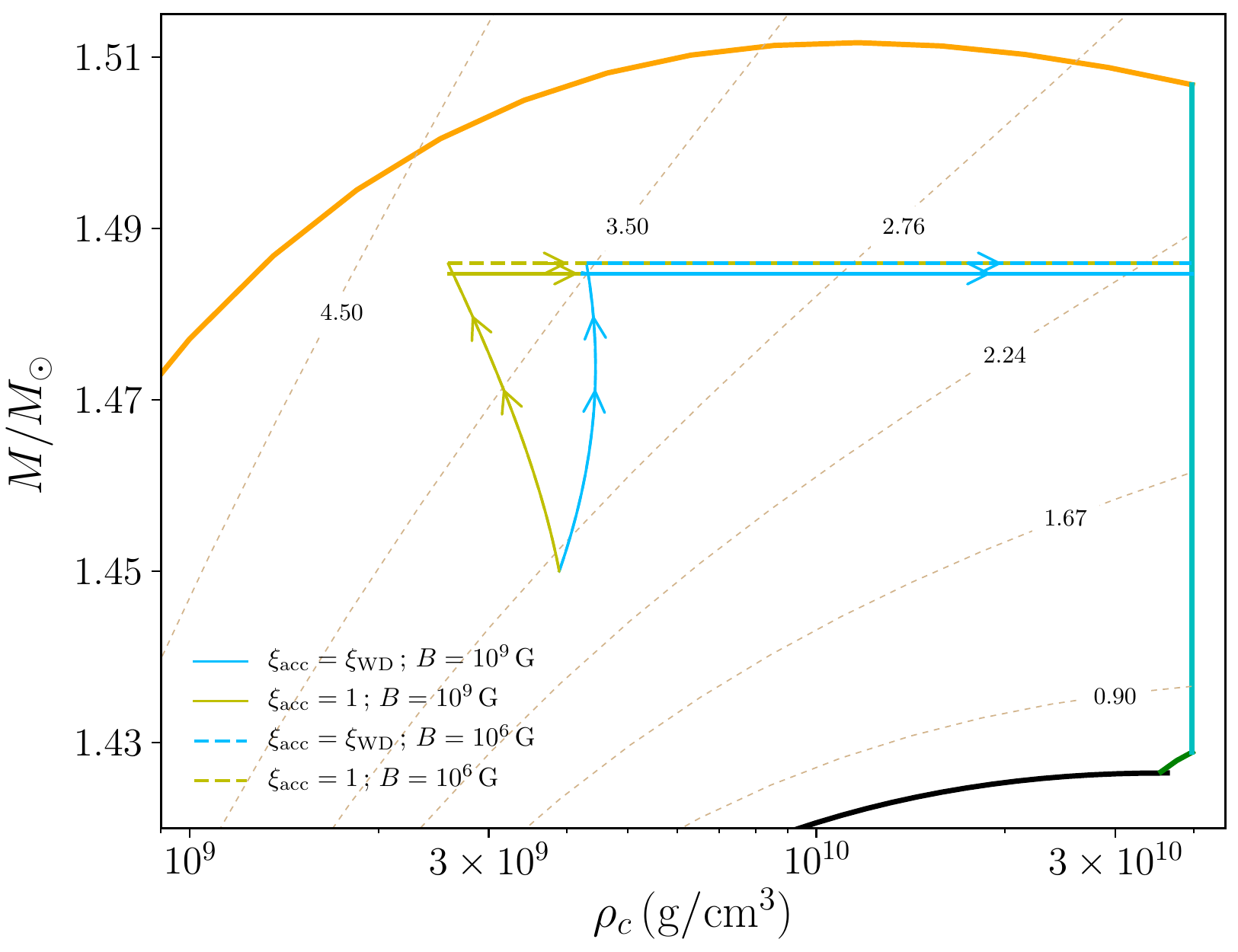}
\caption{Same as  Figure~\ref{fig:WDevol_2}. We compare  models E  (solid lines)
  and G (dashed lines) for two different values of the white dwarf magnetic field, $10^9$~G and $10^{6}$~G, respectively.}
\label{fig:WDevol_B}
\end{figure}

Next, we analyze the influence on the evolution of the strength of the
magnetic field, as we did previously for those models in which
  the accretion rate is computed  using the cooling timescale.  We do
this  comparison adopting  a very  low  value for  the magnetic  field
strength,  $B=10^6$~G  --   model  G  in  Table~\ref{tab:InConModels},
the same value adopted in Section~\ref{sec:tcool}.  The results
of this analysis are shown in Figure~\ref{fig:WDevol_B}.  As can be seen
the  differences  between  the  evolution   of  models  E  and  G  are
minor. Thus, we  conclude that for the typical values  of the magnetic
field  strength originated  in the  merger  of two  white dwarfs,  the
evolution of these models is not significantly affected by the adopted
magnetic field. This can be explained from the fact that, at early times in the viscous timescale prescription, the mass and angular momentum of the white dwarf post-merger evolution is dominated by the accretion torque and the effect of the magnetic field is neglected until the magnetospheric radius equals the white dwarf one. It is in this initial phase that more significant accretion of mass onto the central star occurs. Since, the accretion timescale in this scenarios is so short, the evolution of the magnetospheric radius is dominated by the change of the mass-accretion rate.

\begin{figure}[t]
\centering
\includegraphics[width=0.9\hsize,clip]{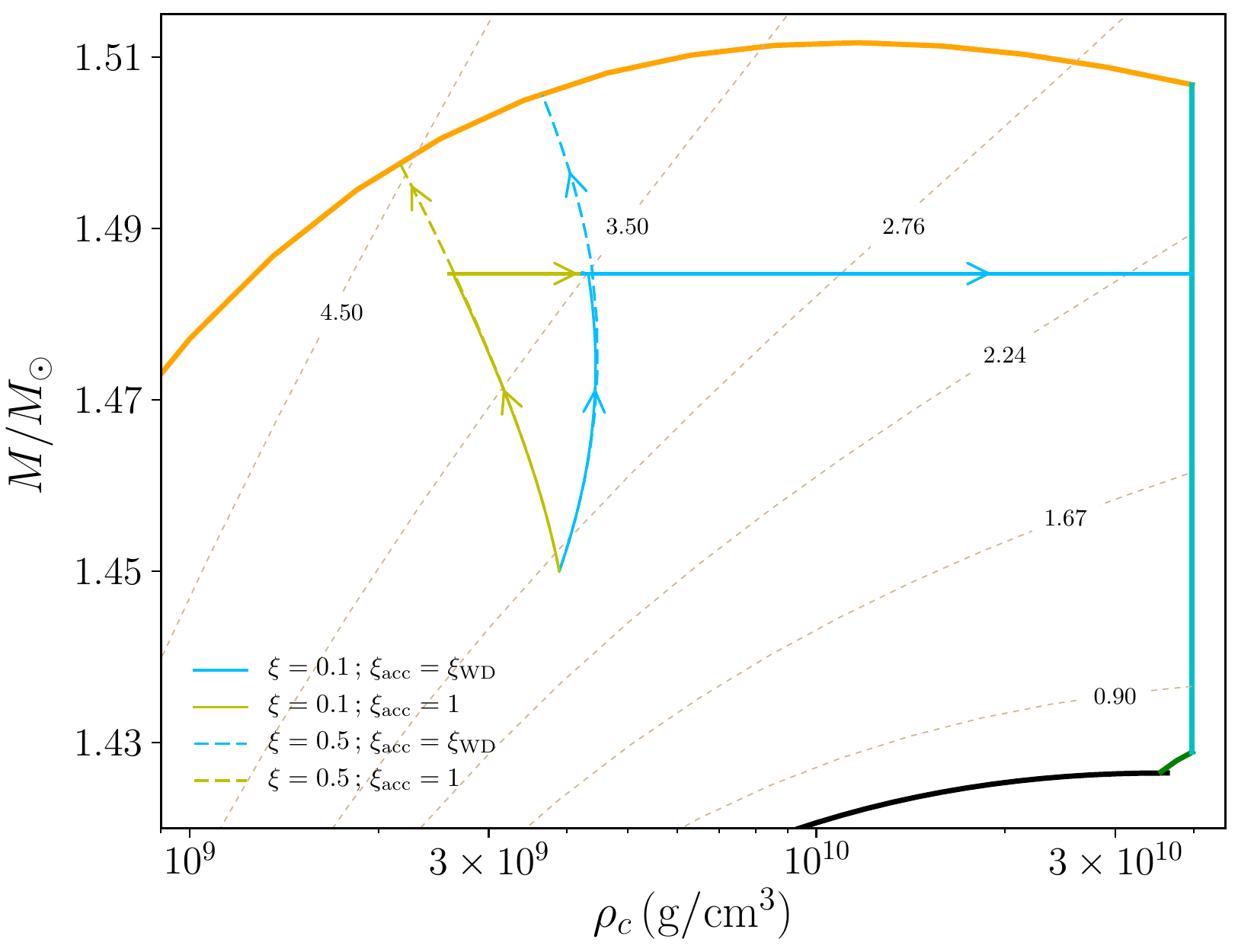}
\caption{Same  as  Figure~\ref{fig:WDevol_2}  for   two  values  of  the
  accretion efficiency parameter,  $\varepsilon=0.1$ (solid lines) and
  0.5 (dashed lines), models E and H, respectively.}
\label{fig:epsilon}
\end{figure}

Another free  parameter of our formulation  is $\varepsilon$,
which  we  recall  measures  how  efficient  accretion  is.   All  the
calculations  presented   until  now  have  been   performed  adopting
$\varepsilon=0.1$. %$ which is a  conservative choice.
However, since the
accreted mass depends significantly on  the precise value of this free
parameter it is important to assess its impact on the results.  We now
study the sensitivity of our calculations to the value adopted for it.
To this end we conducted an  additional set of simulations in which we
adopted     $\varepsilon=0.5$.      This     is     model     H     in
Table~\ref{tab:InConModels}. In Figure~\ref{fig:epsilon}  we compare the
results of  these calculations with those  of model E. As  it could be
expected, this parameter  turns out to be critical,  since it controls
how efficient  accretion is.  Consequently, when  $\varepsilon=0.5$ is
adopted the remnant evolves towards the mass-shedding limit instead of
crossing  the $\beta$-instability  line.   We consider  model  E as  a
reasonable  guess, although  keeping  in mind  that  larger values  of
$\varepsilon$ cannot  be discarded  ``a priori'', and  therefore could
alter the evolution of the remnant.
 
\begin{figure}
\centering
\includegraphics[width=0.9\hsize,clip]{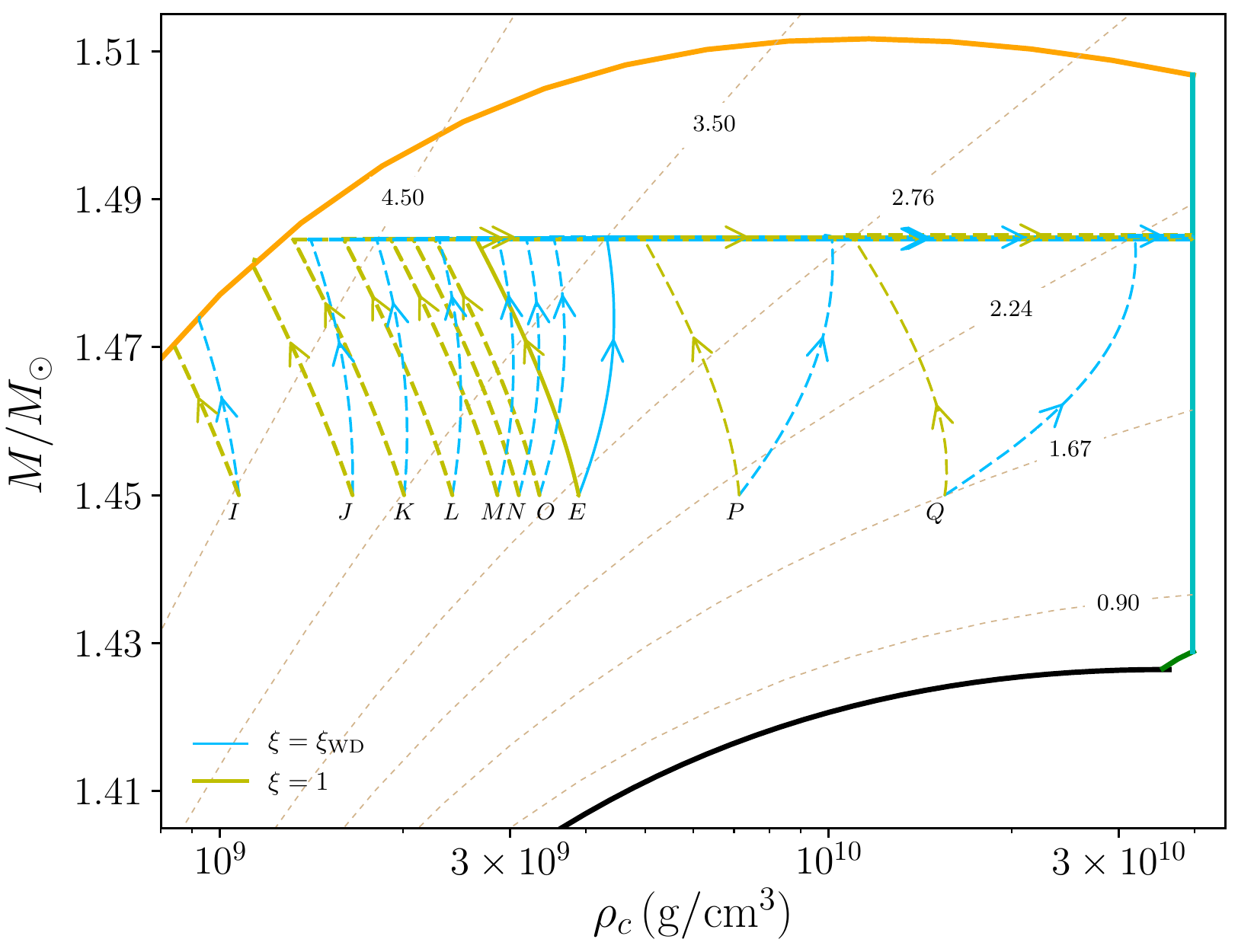}
\caption{Same  as Figure~\ref{fig:WDevol_2}  for several  values of  the
  initial     angular    velocity,     models     I     to    P     in
  Table~\ref{tab:InConModels},  compared to  our  reference case,  for
  which the initial angular  velocity is $\Omega_{\rm WD}=3.00$ (model
  E). From left to right the initial angular velocities of the remnant
  are $\Omega_{\rm WD}=2.30$, $2.50$,  $2.60$, $2.70$, $2.80$, $2.85$,
  $2.90$, $3.50$, and $4.50\, {\rm s}^{-1}$.}
\label{fig:omega}
\end{figure}

Naturally,  another  key ingredient  of  our  approach is  the
  initial angular  velocity of  the central white  dwarf, $\Omega_{\rm
    WD}$.   Figure~\ref{fig:omega} shows  the  evolutionary tracks  for
different values  of the initial  angular velocity of  the post-merger
remnant, keeping unchanged  the rest of initial conditions  of model E
-- models I to Q in  Table~\ref{tab:InConModels} (dashed lines) -- and
we  compare  them with  the  evolutionary  sequence  of model  E,  our
reference model  for this prescription  for the accretion  rate (solid
lines).  As can  be seen, the model with the  smallest initial angular
velocity  --  namely that  with  initial  angular velocity  2.30~${\rm
  rad\,s^{-1}}$, model I -- reaches the mass-shedding limit during the
accretion  phase.  Furthermore,  for  this model  the central  density
decreases,  irrespective of  the  value adopted  for $\xi_{\rm  acc}$.
Model  J, for  which  we adopted  $\Omega_{\rm  WD}=2.50\, {\rm  rad\,
  s}^{-1}$,  only   reaches  the  mass-shedding  limit   if  $\xi_{\rm
  acc}=1.0$, otherwise it reaches the $\beta$ instability region.  The
rest of  the models do  not cross the mass-shedding  instability line.
Note as well that  for model J, as it occurs for  model I, the central
density decreases,  independently of  the value adopted  for $\xi_{\rm
  acc}$.   For model  K  the central  density  decreases if  $\xi_{\rm
  acc}=1.0$ and  increases otherwise.   This is  also true  for models
with even faster rotation rates.

\begin{figure}
\centering
\includegraphics[width=0.9\hsize,clip]{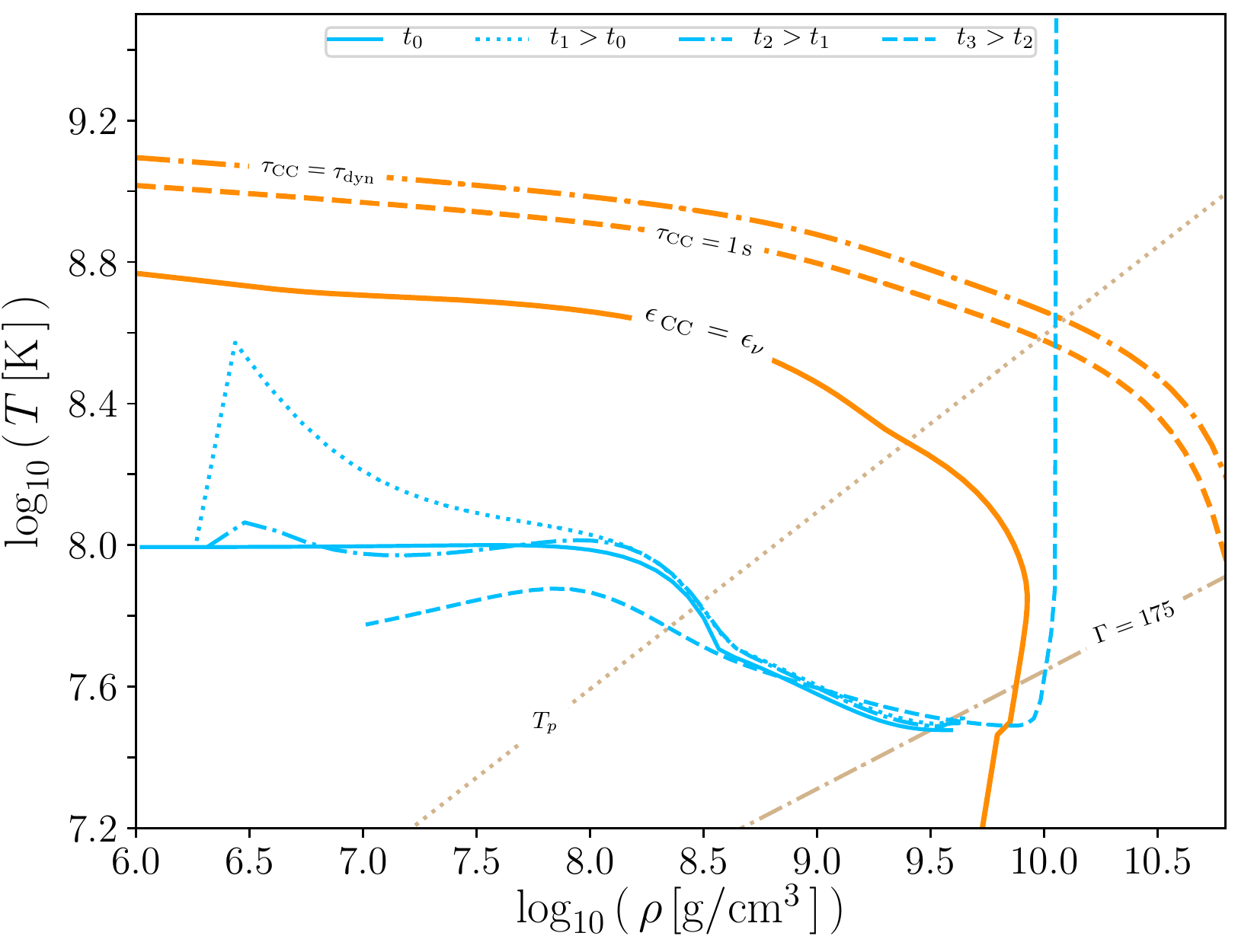}
\caption{Same  as Figure~\ref{fig:WDevol_Trhoprofile}  for  the case  in
  which the initial temperature of model E is smaller. For the sake of
  conciseness we only display the model for which $\xi=1$ is adopted.}
\label{fig:WDevol_cool}
\end{figure}

Additionally,  to  take  into  account  the  effects  of  the  initial
temperature for  the case in  which $\xi=\xi_{\rm acc}$ is  adopted we
computed a model  in which the initial configuration  corresponds to a
white dwarf  in which the  external layers have a  temperature sizably
smaller, namely $10^8$~K,  than that of our standard case  -- that is,
model E --  which was computed adopting a temperature  of the external
layers  $3\times 10^8$~K. The evolution  of this  additional
  model  is shown  in Figure~\ref{fig:WDevol_cool},  and it  is markedly
  different different  from that of  model E.  Specifically,  when the
  adopted temperature  is $10^8$~K, the  external layers of  the white
  dwarf are initially  heated by the accreted material,  as it happens
  also for model E.  However, in this case neutrino cooling (basically
  dominated by neutrino bremsstrahlung) is  so strong that the remnant
  cools rapidly and eventually carbon is  ignited at the center of the
  star, when the central regions have already crystallized.

%%%%%%%%%%%%%%%%%%%%%%%%%%%%%%%%%%%%%%%%%%%%%%%%%
\subsubsection{Evolutionary times}
%%%%%%%%%%%%%%%%%%%%%%%%%%%%%%%%%%%%%%%%%%%%%%%%%

\begin{figure}
\centering
\includegraphics[width=0.9\hsize,clip]{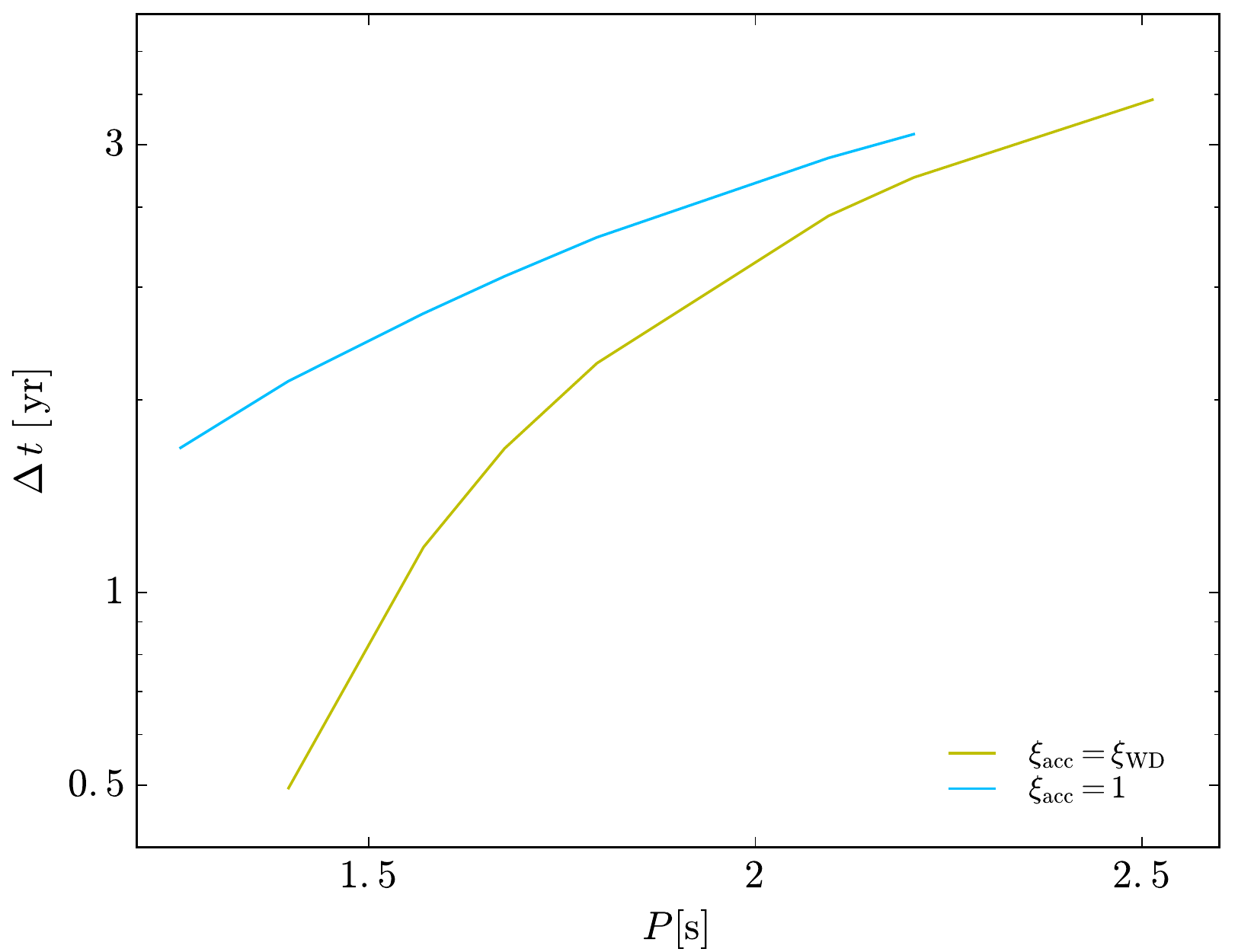}\\
\includegraphics[width=0.9\hsize,clip]{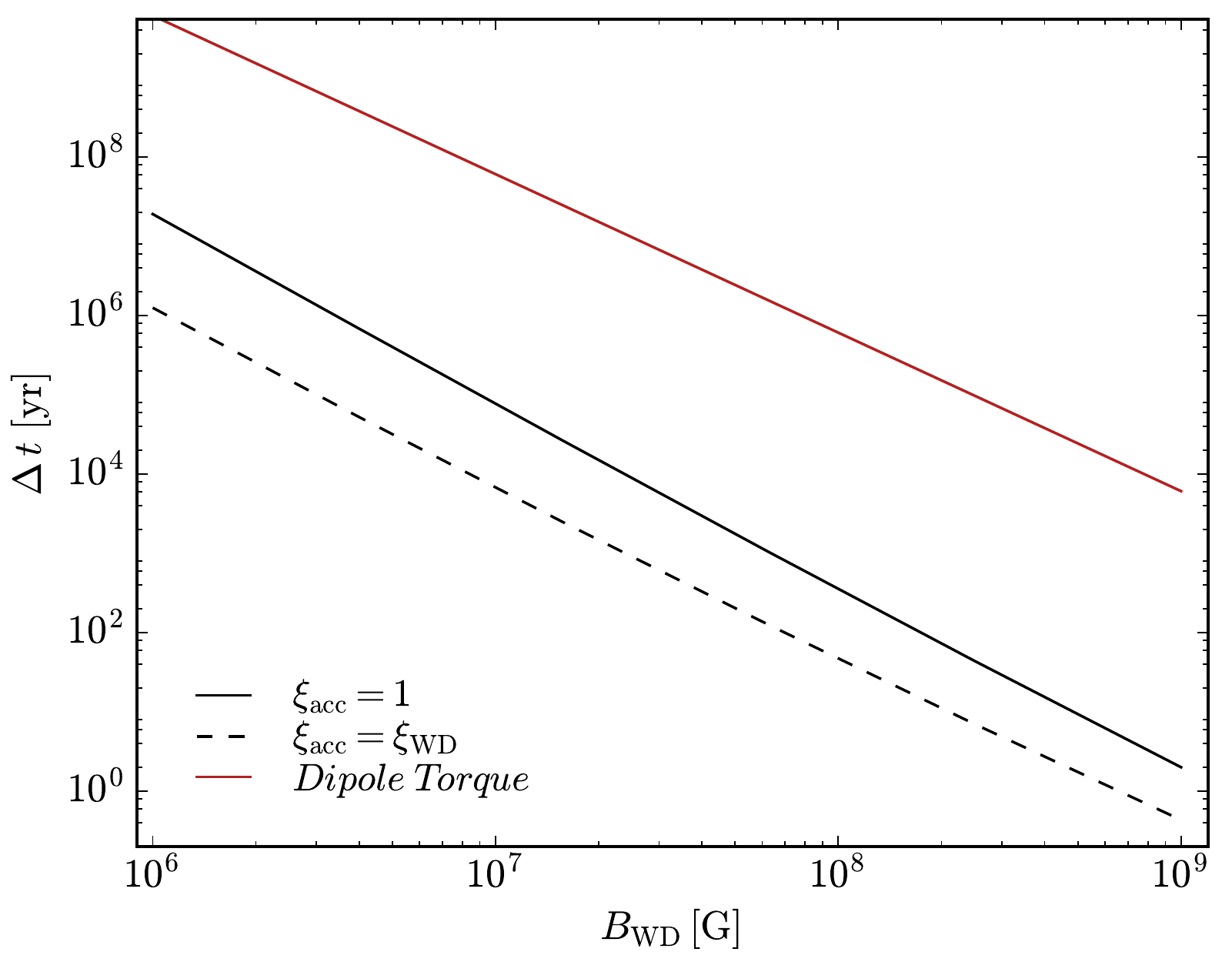}
\caption{Top panel: time necessary to  reach the instability line as a
  function of  of the  initial period of  the remnant.   Bottom panel:
  time  necessary to  reach the  instability  line as  a function  the
  magnetic field of the remnant.}
\label{fig:WDevol_times}
\end{figure}

Finally,  we study  the time  needed to  reach the  instability lines,
$\Delta t$.   We have shown  before that the most  important parameter
that determines the  evolution of the system is the  initial period of
the  system.   In the  top  panel  of Figure~\ref{fig:WDevol_times}  the
dependence of $\Delta  t$ on the initial period of  the rotating white
dwarf resulting from the merger is  displayed, for our two choices for
the value of  $\xi_{\rm acc}$. Clearly, the larger the  period is, the
longer $\Delta t$  is. This is the natural consequence  of the smaller
initial centrifugal force.

The second  important parameter that  determines the duration  of this
evolutionary  phase is  the  strength  of the  magnetic  field of  the
remnant.  We showed  before that the value of $B_{\rm  WD}$ has little
effect   on   the  outcome   of   the   post-merger  system   --   see
Figure~\ref{fig:WDevol_B}.   Notwithstanding,  $B_{\rm  WD}$  influences
considerably $\Delta t$.  This occurs  because the larger $B_{\rm WD}$
is, the  larger is  the magnetospheric radius.   Thus, we  expect that
large values  of $B_{\rm  WD}$ would  result in  substantially smaller
values of $\Delta t$. To illustrate  this in a quantitative manner, in
the bottom panel of  Figure~\ref{fig:WDevol_times} the total time needed
to  reach the  instability  region is  plotted as  a  function of  the
surface magnetic field of the remnant.   As can be seen, the evolution
of the  systems is indeed  faster for large magnetic  field strengths.
This behavior is natural since  for large magnetic field strengths the
central remnant is more tightly coupled with the surrounding Keplerian
disk.  Consequently, the accretion phase  is shorter.  In addition, in
this panel we also show (using a solid, red line) the time it takes to
the remnant  to become unstable  when only magnetic dipole  braking is
considered. Since in this case, there is no accretion, the star evolves in a constant-mass sequence. From Equation~(\ref{eq:Tdipole}), we obtained:
\begin{equation}\label{eq:delta_B}
\Delta\, t = \frac{c^3}{2 B_{\rm WD}^2}\int_{J_{{\rm WD},0}}^{J_{{\rm WD},f}} \frac{dJ_{\rm WD}}{\Omega_{\rm WD}^3R_{\rm WD}^6}
\end{equation}
It can be seen that $\Delta\, t$ is inversely proportional to the square of the magnetic field. \citet{2012MNRAS.419.1695I} gives an analytical expression of Equation~(\ref{eq:delta_B}) in the approximate case when changes in the white dwarf structure due to the configuration rearrangement during the angular momentum loss are neglected). It is important to mention that such changes are very important for the determination of the rotational evolution of super-Chandrasekhar white dwarfs \citep{2013ApJ...762..117B}. Clearly, the evolution of the system is always much faster when all  the torques acting on  the remnant are correctly  taken into
account.  Furthermore, it  is important to realize  that the evolution
is even faster when matter is accreted onto the surface of the remnant
with  the  Keplerian  velocity.   Finally,  note  as  well  that  when
$\xi_{\rm  acc}=1.0$ is  adopted, moderately  longer durations  of the
post-merger phase, when  compared with the case in  which the accreted
matter has the  Keplerian velocity at the surface of  the remnant, are
obtained.

%%%%%%%%%%%%%%%%%%%%%%%%%%%%%%%%%%%%%%%%%%%%%%%%%
%%%%%%%%%%%%%%%%%%%%%%%%%%%%%%%%%%%%%%%%%%%%%%%%%
\section{Comparison with previous works}
\label{sec:comparison}
%%%%%%%%%%%%%%%%%%%%%%%%%%%%%%%%%%%%%%%%%%%%%%%%%
%%%%%%%%%%%%%%%%%%%%%%%%%%%%%%%%%%%%%%%%%%%%%%%%%
\begin{table*}[t]
\centering
\caption{Comparison with previous works}
\renewcommand{\arraystretch}{1.5}
\begin{tabular}{C{2.5cm}C{2.15cm}m{3.5cm}cC{2.5cm}C{2.0cm}}
  \hline
  \hline
  Work &  Merger Configuration & \centering Post-merger configurations & Magnetic Field & Magnetic field configuration& Carbon Ignition  \\
  \hline
  \citet{2007MNRAS.380..933Y} & $0.9\, M_\odot - 0.6\, M_\odot$ 
$0.9\, M_\odot - 0.7\, M_\odot$  
$1.0\, M_\odot - 1.0\, M_\odot$ 
& slowly rotating cold core ($0.6\, M_\odot$) with a rapidly rotating hot envelope ($0.5\,M_\odot$) + a Keplerian disc ($0.4\, M_\odot$) & - & - & Center/off-center ignition \\ \hline
 \citet{vk10}& $0.6\, M_\odot  - 0.6\, M_\odot$ C/O WD& rapidly rotating WD + thick disk & - & - & Center ignition \\ \hline
 \citet{2012ApJ...748...35S} & $0.6\, M_\odot - 0.9\, M_\odot$ C/O WD & WD ($\sim 0.9\, M_\odot$) with a hot, slowly rotating and radially extended envelope supported by thermal pressure ($\sim 0.6\, M_\odot$)& - & - & Off-center ignition that lead a high mass O/Ne WD\\ \hline
 \citet{2012MNRAS.427..190S} / \citet{2016MNRAS.463.3461S} & $0.6\, M_\odot -0.9\, M_\odot$ $0.9\, M_\odot - 1.2\, M_\odot$ C/O WD & WD with a thermally supported envelope ($1.5\,M_\odot$) & - & - & Off-center ignition that leads an O/Ne WD\\ \hline
 \citet{Ji} & $0.6\, M_\odot - 0.6\, M_\odot$ C/O WD&  rotating WD ($0.96\, M_\odot$) surrounding by a hot corona ($0.04\, M_\odot$) + thick disk ($0.2\, M_\odot$) & $2\times 10^8$~G & $\cfrac{B_t}{B_p}\sim 1.5$ & Center Ignition \\ \hline
  \citet{2015ApJ...806L...1Z} & $0.625\, M_\odot$-$0.65\, M_\odot$ C/O WD& WD  ($0.64\, M_\odot$) surrounded by thermally supported hot envelope  ($0.42 \, M_\odot$) + disk ($0.21\, M_\odot$) &$4 \times 10^{10}$~G& $\cfrac{E_{Bt}}{E_{B}}=0.6 $ & - \\ \hline
  This work & $0.78\, M_\odot$-$1.12\, M_\odot$ C/O WD& $1.45 \, M_\odot$ rigidly rotating super-Chandrasekhar WD + $0.45\,M_\odot$ thin Keplerian disk & $[10^{6}, 10^{9}]$~G& Poloidal &  Center/off center ignition \\
 \hline
 \hline
\end{tabular}
\tablecomments{$B_t/B_p:$ ratio of the toroidal to the poloidal field strength. $E_{Bt}/E_{B}:$ ratio of the toroidal to the total magnetic energy. }
\label{tab:Models_comparison}
\end{table*}
%Clearly our semi-analytical approach allow a more efficient exploration of the parameters space that a full magnetohydrodynamics simulations.

A detailed study of the evolution of the remnant of white dwarf mergers started with the work of \citet{2007MNRAS.380..933Y}. They  mapped the final configuration of the SPH simulations of the merger of a $0.9$--$0.6\,M_\odot$ binary into a 1-D stellar evolution hydrodynamic code and follow its forward evolution. Their merger remnant is represented by a cold and slowly rotating rigid core surrounding by a hot and rapid rotating envelope with a Keplerian disk around it. They allowed the accretion from the disk by a constant  accretion rate (of the order of the Eddington limit) and include a prescription for  the transport of angular momentum with  timescales of the order of the thermal timescale. They conclude that a off-center carbon ignition can be avoided if the transport angular momentum timescale is greater that the cooling timescale by neutrino emission and the accretion rate on the star is slow enough, i,e, $\dot{M}_{\rm WD}<5\times 10^{-6}$--$10^{-5}\,M_\odot$~yr$^{-1}$. We improved this approach allowing the mass-accretion rate to vary with time, consistently with the white dwarf thermal evolution. Additionally, we introduced a  framework for evolution of the angular momentum of the post-merger white dwarf including the torque that acts on the star taking into account the magnetic field effect of the central white dwarf on the evolution of the post-merger configuration. We have found that strong magnetic fields, $B_{\rm WD}>10^7$~G can also avoid an off-center carbon ignition.

\citet{vk10} estimated the post-merger evolution based on the results of the SPH simulations of \citep{2009A&A...500.1193L} and concluded that the accretion occurs in a shorter timescale ($\sim 10$~h), causinga compression of the white dwarf core, with a consequent increase of its temperature until it reaches the carbon-ignition runaway leading to a delayed explosion. However, they focus on mass-symmetric white dwarf mergers ($0.6$--$0.6\,M_\odot$) in which the two white dwarfs are disrupting and the final remnant configuration has a temperature profile that peaks at the center.

A different approach was presented in \citet{2012ApJ...748...35S} and in \citet{2012MNRAS.427..190S,2016MNRAS.463.3461S}.  In
  these works it was argued that,  due to the differential rotation of
  the  system,   after  the   dynamical  phase   of  the   merger  the
  magneto-rotational  instability  becomes  effective, and  that  in  a
  viscous timescale $~10^4$~s (orders  of magnitude shorter than the
  thermal  timescale $\sim  10^4$~yr) the angular  momentum  of  the
  tidally disrupted white dwarf can be redistributed in the surface of
  the central  star, leading  to a rotating  configuration with  a hot
  envelope   and with    almost   all   the   mass    of   the   secondary
  star.  These works also computed the thermal  evolution of the post-merger remnant characterized with a timescale of the order of $10^3$--$10^4$~yr. \citet{2012MNRAS.427..190S} studied the merger of a $0.9-0.6\, M_\odot$ white dwarf binary, and \cite{2012MNRAS.427..190S,2016MNRAS.463.3461S} extended the initial conditions parameter space but the  evolution of  post-merger configuration in presence of a magnetic  field of the central white dwarf was not there considered. These configurations spin down during the viscous evolution, while our super-Chandrasekhar remnants spin up during all the evolution, even in the case of angular momentum losses.
 
There are two works considering the effect of the magnetic field during the dynamical timescale of the merger \citep{2015ApJ...806L...1Z}  and in the viscous timescale of the post-merger evolution \citep{Ji}. In these works, the evolution of the system magnetic field is followed in a different fashion with respect to our approach, since we have adopted a dipole magnetic field with constant magnitude and inclination angle. Assuming as initial condition the remnant of the merger of a $0.6$--$0.6\,M_\odot$ carbon-oxygen white dwarf binary from the SPH simulation of \citet{ 2009A&A...500.1193L}, \citet{Ji} evolves the system for $3\times 10^4$~s with the  {\scshape Flash} code in a 2D axisymmetric cylindrical Eulerian grid. They introduced a weak poloidal magnetic field and showed that the magneto-rotational instability developed in the disk leads to a rapid growth of its  magnetic field, the spin down of the white dwarf remnant and its magnetization to field strengths around $\sim 2\times 10^8$~G.  They computed an effective magnetic Shakura-Sunyaev parameter of the order of $\langle \alpha_m\rangle\sim 0.01$, a value one order of magnitude smaller than the $\alpha$ we adopted here.  They found that the white dwarf magnetic field varies with time, indicating a disordered interior magnetic field.  At the end of the simulation the magnetic field toroidal component is $1.5$ times bigger than the poloidal one. In our model we have assumed a dipole magnetic field aligned with the white dwarf rotation axis. In this case, the magnetic field, has no toroidal component. However, the simulations of \citet{Ji} have limited resolution and the field strength are affected by numerical resistivity. In addition, in these magneto-hydrodynamics simulations the disk lost almost 90\% of its initial mass,  with 82\% of it accreted by the white dwarf remnant and the rest going into outflows of which about $10^{-3}M_\odot$ are ejected and unbound to the system. The central white dwarf spins down, losing angular momentum due to the development of Maxwell stresses at the white dwarf boundary. A directed comparison is difficult, since we have studied super-Chandrasekhar white dwarf with initial angular velocity one order of magnitude higher than the  one studied by \citet{Ji}  ($\Omega_{\rm WD,0}=0.03$~s$^{-1}$). However, one difference is that in our model the white dwarf first gain angular momentum due to the mass accretion, and then due to the magnetic torque loses it.

\citet{2015ApJ...806L...1Z} simulated the merger of a $0.625$--$0.65\, M_\odot$ carbon-oxygen white dwarf binary giving to each white dwarf a dipole seed magnetic field with the moving-mesh code {\scshape Arepo} \citep{2010MNRAS.401..791S}. They found that during the merger dynamics, the magnetic field were too weak to have an effect in the evolution, obtaining a final remnant composed by a degenerate core with a thermally supported envelope surrounding by a rotationally supported disk. This configuration is similar to the one obtained in the SPH simulations \citep{2013ApJ...767..164Z}. However, the remnant magnetic field have a complex structure with a a volume average field strength $>10^{10}$~G in the core, with both poloidal and toroidal components.
%This is in contrast with the \citet{Ji} initial seed magnetic field, a probably change in the merger evolution.}

In Table~\ref{tab:Models_comparison} is summarized the main features of the post-merger evolution studied in this paper and the comparison with the ones of the works discussed above. We specified the binary mass of the merger configurations  studied, the post-merger configurations after the merger (for the works of \cite{2007MNRAS.380..933Y,vk10,2013ApJ...767..164Z}) or after the viscous evolutions of the system (for the work of \cite{2012ApJ...748...35S,2012MNRAS.427..190S,Ji}), the magnitude and configuration of the central remnant magnetic field (if it is considered). In the last column, we specified  if the central remnant  developed  conditions for an off center or center carbon ignition.

%%%%%%%%%%%%%%%%%%%%%%%%%%%%%%%%%%%%%%%%%%%%%%%%%
%%%%%%%%%%%%%%%%%%%%%%%%%%%%%%%%%%%%%%%%%%%%%%%%%
\section{Conclusions}
\label{sec:concl}
%%%%%%%%%%%%%%%%%%%%%%%%%%%%%%%%%%%%%%%%%%%%%%%%%
%%%%%%%%%%%%%%%%%%%%%%%%%%%%%%%%%%%%%%%%%%%%%%%%%

The evolution of  the remnant of the merger of  a binary white
  dwarf is still an  open problem. Detailed hydrodynamical simulations
  show that  the product  of the  merger consists  of a  central white
  dwarf that  rotates as a  rigid body,  surrounded by a  hot, rapidly
  rotating corona --  which has been proven to  produce large magnetic
  fields  -- and  a  Keplerian disk.   In this  paper  we studied  the
  evolution of metastable, magnetized super-Chandrasekhar white dwarfs
  formed in the  aftermath of the merger of close  binary systems made
  of two white dwarfs.

Our simulations take into  account the magnetic torques acting
  on the star, accretion from the Keplerian disk, and the threading of
  the magnetic  field lines  through the  disk --  therefore improving
  previous   calculations  of   this   kind.    Furthermore,  in   our
  computations  --  at odds  with  previous  efforts  -- we  employ  a
  mass-radius  relationship  for  rotating  white  dwarfs.  Also,  our
  calculations incorporate  the thermal evolution of  the white dwarf. 
  %All in all,  the set of calculations presented here  relies on solid physical grounds.

Furthermore, our  simulations   were  performed   using  two
  different prescriptions for  the mass accretion rate  on the central
  white dwarf. In  a first set of simulations we  adopted an accretion
  rate set by cooling timescale of  the Keplerian disk, whereas in the
  second suite of models the adopted accretion rate was computed using
  the viscous timescale  of the disk. These two  prescriptions for the
  accretion rate  cover a  very large range  of values,  and therefore
  allow us  to investigate the  possible outcomes of the  evolution of
  these systems in a quite generic way.
 
  Finally, we  also explored the  effects of the adopted  set of
  initial parameters. These include the  mass of the remnant star, its
  radius,  angular velocity,  and moment  of inertia,  as well  as the
  strength of  the magnetic  field.  We showed  that the  timescale in
  which  the   newly  formed   white  dwarf   evolves  to   reach  the
  thermodynamical conditions  for carbon be burned  explosively, or to
  reach the Keplerian  mass-shedding, secular axisymmetric instability
  or inverse  beta decay  instability depends  crucially on  all these
  parameters.
  
However, we have made a number of assumptions that might be  relaxed in forthcoming works. For instance, to allow a non-zero inclination angle between the spin axis and the orientation of the dipole magnetic field and its evolution with time. Also, the post-merger central remnant rotates differentially between the core and the corona, while we have adopted a totally rigid central remnant. Thus the model can be improved by implementing a transport mechanism for the angular momentum in the interior of the central white dwarf remnant.

For the accretion rate prescription set by the viscous time scale, we have adopted a geometrically thin disk model.  At the early times in the evolution of the post-merger remnant,  we obtain highly super-Eddington accretion rate of up to $10^{-1}\, M\odot$~s$^{-1}$. The dissipation required to produce the needed very short viscous timescales might heat the disk to the point of carbon ignition \citep{1990A&A...236..378M}, but we did not consider these heating effects in the  present work. The disk of the post-merger remnant could be best described by a thick disk model. This might be also study in a forthcoming work.

On the other hand, we have approximated the white dwarf magnetic field to a prefect dipole  configuration but this is not necessary the case as it has been shown by the MHD simulations of \citet{Ji} and \citet{2015ApJ...806L...1Z}. This work could be extended in order to introduce additional configurations of the magnetic field and its interaction with the surrounding disk.
  
We  showed that  in most  of our  models carbon  reactions are  highly
efficient  in heating  the interior  of the  remnant, with  timescales
shorter than  the dynamical time.  This  can happen both in  the outer
layers  or at  the center  of the  newly formed  white dwarf,
depending on  the initial  conditions of  the white  dwarf and  on the
efficiency  of  mass accretion  and  angular  momentum evolution  (see
Figure~\ref{fig:WDevol_Trhoprofile}). In most  cases, the time
it takes to the star to reach explosive conditions is shorter than the
one needed to reach the  inverse beta-decay instability or the secular
instability, which lead to gravitational collapse.  Hence, we conclude
that  the central  white  dwarf  can reach  conditions  for a  delayed
explosion for a sufficiently  broad set of initial conditions.
  Whether  carbon is  ignited at  the center  of is  burned off-center
  depends crucially  on the  magnetic field strength and is almost independently of the post-merger WD initial angular velocity.  Our simulations
  show  that  when  the  magnetic  field is  weak  carbon  is  ignited
  off-center, while central  explosions are the outcome  when a ordered global strong
  dipole magnetic field is  produced in the hot,  rapidly rotating convective
  corona formed  in the aftermath  of white dwarf  mergers. Naturally,
  this depends on whether the stellar dynamo is saturated or not.

In  summary, spinning,  magnetized, super-Chandrasekhar  white dwarfs,
resulting from the merger of two less massive white dwarfs that do not
produce a Type Ia supernova in  a violent merger can produce a delayed
explosion, provided  that the remnant  is massive enough and  a strong
magnetic field is produced during the merger.

\acknowledgments  During the refereeing process of this work our colleague but overall dear friend, Enrique Garc\'ia-Berro, sadly passed away. To him and to his memory, we dedicate this work and express all our gratitude. We thank the Referee for the detailed comments and suggestions which help to improve the presentation of our results. Part of  this  work was  supported  by MINECO  grant AYA2014-59084-P, and by the AGAUR.

%\bibliographystyle{aasjournal}
%\bibliography{spin}

\begin{thebibliography}{}
\expandafter\ifx\csname natexlab\endcsname\relax\def\natexlab#1{#1}\fi

\bibitem[{{Aznar-Sigu{\'a}n} {et~al.}(2013){Aznar-Sigu{\'a}n},
  {Garc{\'{\i}}a-Berro}, {Lor{\'e}n-Aguilar}, {Jos{\'e}}, \&
  {Isern}}]{Aznarsiguan2013}
{Aznar-Sigu{\'a}n}, G., {Garc{\'{\i}}a-Berro}, E., {Lor{\'e}n-Aguilar}, P.,
  {Jos{\'e}}, J., \& {Isern}, J. 2013, \mnras, 434, 2539

\bibitem[{{Aznar-Sigu{\'a}n} {et~al.}(2015){Aznar-Sigu{\'a}n},
  {Garc{\'{\i}}a-Berro}, {Lor{\'e}n-Aguilar}, {Soker}, \&
  {Kashi}}]{aznarsiguanetal15}
{Aznar-Sigu{\'a}n}, G., {Garc{\'{\i}}a-Berro}, E., {Lor{\'e}n-Aguilar}, P.,
  {Soker}, N., \& {Kashi}, A. 2015, \mnras, 450, 2948

\bibitem[{{Beloborodov}(2014)}]{Beloborodov14}
{Beloborodov}, A.~M. 2014, \mnras, 438, 169

\bibitem[{{Benz} {et~al.}(1990){Benz}, {Cameron}, {Press}, \&
  {Bowers}}]{1990ApJ...348..647B}
{Benz}, W., {Cameron}, A.~G.~W., {Press}, W.~H., \& {Bowers}, R.~L. 1990, \apj,
  348, 647

\bibitem[{{Boshkayev} {et~al.}(2013){Boshkayev}, {Rueda}, {Ruffini}, \&
  {Siutsou}}]{2013ApJ...762..117B}
{Boshkayev}, K., {Rueda}, J.~A., {Ruffini}, R., \& {Siutsou}, I. 2013, \apj,
  762, 117

\bibitem[{{Cannizzo} {et~al.}(1990){Cannizzo}, {Lee}, \&
  {Goodman}}]{1990ApJ...351...38C}
{Cannizzo}, J.~K., {Lee}, H.~M., \& {Goodman}, J. 1990, ApJ, 351, 38

\bibitem[{{Chabrier} \& {Potekhin}(1998)}]{1998PhRvE..58.4941C}
{Chabrier}, G., \& {Potekhin}, A.~Y. 1998, \pre, 58, 4941

\bibitem[{{Chandrasekhar}(1931)}]{1931ApJ....74...81C}
{Chandrasekhar}, S. 1931, \apj, 74, 81

\bibitem[{{Chandrasekhar}(1970)}]{1970ApJ...161..571C}
---. 1970, \apj, 161, 571

\bibitem[{{Chatterjee} {et~al.}(2000){Chatterjee}, {Hernquist}, \&
  {Narayan}}]{2000ApJ...534..373C}
{Chatterjee}, P., {Hernquist}, L., \& {Narayan}, R. 2000, \apj, 534, 373

\bibitem[{{Dan} {et~al.}(2014){Dan}, {Rosswog}, {Br{\"u}ggen}, \&
  {Podsiadlowski}}]{2014MNRAS.438...14D}
{Dan}, M., {Rosswog}, S., {Br{\"u}ggen}, M., \& {Podsiadlowski}, P. 2014,
  \mnras, 438, 14

\bibitem[{{Dan} {et~al.}(2011){Dan}, {Rosswog}, {Guillochon}, \&
  {Ramirez-Ruiz}}]{2011ApJ...737...89D}
{Dan}, M., {Rosswog}, S., {Guillochon}, J., \& {Ramirez-Ruiz}, E. 2011, \apj,
  737, 89

\bibitem[{{Davidson} \& {Ostriker}(1973)}]{1973ApJ...179..585D}
{Davidson}, K., \& {Ostriker}, J.~P. 1973, \apj, 179, 585

\bibitem[{{Eggleton}(1983)}]{1983ApJ...268..368E}
{Eggleton}, P.~P. 1983, \apj, 268, 368

\bibitem[{{Ertan} {et~al.}(2009){Ertan}, {Ek{\c s}i}, {Erkut}, \&
  {Alpar}}]{2009ApJ...702.1309E}
{Ertan}, {\"U}., {Ek{\c s}i}, K.~Y., {Erkut}, M.~H., \& {Alpar}, M.~A. 2009,
  \apj, 702, 1309

\bibitem[{{Friedman} {et~al.}(1988){Friedman}, {Ipser}, \&
  {Sorkin}}]{1988ApJ...325..722F}
{Friedman}, J.~L., {Ipser}, J.~R., \& {Sorkin}, R.~D. 1988, \apj, 325, 722

\bibitem[{{Garc{\'{\i}}a-Berro} {et~al.}(2012){Garc{\'{\i}}a-Berro},
  {Lor{\'e}n-Aguilar}, {Aznar-Sigu{\'a}n}, {Torres}, {Camacho}, {Althaus},
  {C{\'o}rsico}, {K{\"u}lebi}, \& {Isern}}]{2012ApJ...749...25G}
{Garc{\'{\i}}a-Berro}, E., {Lor{\'e}n-Aguilar}, P., {Aznar-Sigu{\'a}n}, G.,
  {et~al.} 2012, \apj, 749, 25

\bibitem[{{Gasques} {et~al.}(2005){Gasques}, {Afanasjev}, {Aguilera}, {Beard},
  {Chamon}, {Ring}, {Wiescher}, \& {Yakovlev}}]{2005PhRvC..72b5806G}
{Gasques}, L.~R., {Afanasjev}, A.~V., {Aguilera}, E.~F., {et~al.} 2005, \prc,
  72, 025806

\bibitem[{{Ghosh} \& {Lamb}(1979)}]{1979ApJ...232..259G}
{Ghosh}, P., \& {Lamb}, F.~K. 1979, \apj, 232, 259

\bibitem[{{Goldreich} \& {Julian}(1969)}]{1969ApJ...157..869G}
{Goldreich}, P., \& {Julian}, W.~H. 1969, \apj, 157, 869

\bibitem[{{Han} \& {Podsiadlowski}(2004)}]{Han2004}
{Han}, Z., \& {Podsiadlowski}, P. 2004, \mnras, 350, 1301

\bibitem[{{Hartle}(1967)}]{1967ApJ...150.1005H}
{Hartle}, J.~B. 1967, \apj, 150, 1005

\bibitem[{{Hartle} \& {Thorne}(1968)}]{1968ApJ...153..807H}
{Hartle}, J.~B., \& {Thorne}, K.~S. 1968, \apj, 153, 807

\bibitem[{{Henyey} \& {L'Ecuyer}(1969)}]{1969ApJ...156..549H}
{Henyey}, L., \& {L'Ecuyer}, J. 1969, \apj, 156, 549

\bibitem[{{Iben} \& {Tutukov}(1984)}]{IT84}
{Iben}, Jr., I., \& {Tutukov}, A.~V. 1984, \apjs, 54, 335

\bibitem[{{Ilkov} \& {Soker}(2013)}]{Ilkov}
{Ilkov}, M., \& {Soker}, N. 2013, \mnras, 428, 579

\bibitem[{{Illarionov} \& {Sunyaev}(1975)}]{1975A&A....39..185I}
{Illarionov}, A.~F., \& {Sunyaev}, R.~A. 1975, \aap, 39, 185

\bibitem[{{Itoh} {et~al.}(1996){Itoh}, {Hayashi}, {Nishikawa}, \&
  {Kohyama}}]{1996ApJS..102..411I}
{Itoh}, N., {Hayashi}, H., {Nishikawa}, A., \& {Kohyama}, Y. 1996, \apjs, 102,
  411

\bibitem[{{Itoh} {et~al.}(1984){Itoh}, {Kohyama}, {Matsumoto}, \&
  {Seki}}]{1984ApJ...285..758I}
{Itoh}, N., {Kohyama}, Y., {Matsumoto}, N., \& {Seki}, M. 1984, \apj, 285, 758

\bibitem[{{Itoh} {et~al.}(1983){Itoh}, {Mitake}, {Iyetomi}, \&
  {Ichimaru}}]{1983ApJ...273..774I}
{Itoh}, N., {Mitake}, S., {Iyetomi}, H., \& {Ichimaru}, S. 1983, \apj, 273, 774

\bibitem[{{Ji} {et~al.}(2013){Ji}, {Fisher}, {Garc{\'{\i}}a-Berro},
  {Tzeferacos}, {Jordan}, {Lee}, {Lor{\'e}n-Aguilar}, {Cremer}, \&
  {Behrends}}]{Ji}
{Ji}, S., {Fisher}, R.~T., {Garc{\'{\i}}a-Berro}, E., {et~al.} 2013, \apj, 773,
  136

\bibitem[{{Kashi} \& {Soker}(2011)}]{Soker}
{Kashi}, A., \& {Soker}, N. 2011, \mnras, 417, 1466

\bibitem[{{K{\"u}lebi} {et~al.}(2013){K{\"u}lebi}, {Ek{\c s}i},
  {Lor{\'e}n-Aguilar}, {Isern}, \& {Garc{\'{\i}}a-Berro}}]{kulebi}
{K{\"u}lebi}, B., {Ek{\c s}i}, K.~Y., {Lor{\'e}n-Aguilar}, P., {Isern}, J., \&
  {Garc{\'{\i}}a-Berro}, E. 2013, \mnras, 431, 2778

\bibitem[{{Leonard}(2007)}]{Leonard}
{Leonard}, D.~C. 2007, \apj, 670, 1275

\bibitem[{{Livio} \& {Riess}(2003)}]{Livio}
{Livio}, M., \& {Riess}, A.~G. 2003, \apjl, 594, L93

\bibitem[{{Lor{\'e}n-Aguilar} {et~al.}(2009){Lor{\'e}n-Aguilar}, {Isern}, \&
  {Garc{\'{\i}}a-Berro}}]{2009A&A...500.1193L}
{Lor{\'e}n-Aguilar}, P., {Isern}, J., \& {Garc{\'{\i}}a-Berro}, E. 2009, \aap,
  500, 1193

\bibitem[{{Menou} {et~al.}(1999){Menou}, {Esin}, {Narayan}, {Garcia}, {Lasota},
  \& {McClintock}}]{1999ApJ...520..276M}
{Menou}, K., {Esin}, A.~A., {Narayan}, R., {et~al.} 1999, \apj, 520, 276

\bibitem[{{Nomoto}(1982)}]{1982ApJ...253..798N}
{Nomoto}, K. 1982, \apj, 253, 798

\bibitem[{{Nomoto} \& {Iben}(1985)}]{1985ApJ...297..531N}
{Nomoto}, K., \& {Iben}, Jr., I. 1985, \apj, 297, 531

\bibitem[{{Pakmor} {et~al.}(2010){Pakmor}, {Kromer}, {R{\"o}pke}, {Sim},
  {Ruiter}, \& {Hillebrandt}}]{pakmor}
{Pakmor}, R., {Kromer}, M., {R{\"o}pke}, F.~K., {et~al.} 2010, \nat, 463, 61

\bibitem[{{Perlmutter} {et~al.}(1999){Perlmutter}, {Aldering}, {Goldhaber},
  {Knop}, {Nugent}, {Castro}, {Deustua}, {Fabbro}, {Goobar}, {Groom}, {Hook},
  {Kim}, {Kim}, {Lee}, {Nunes}, {Pain}, {Pennypacker}, {Quimby}, {Lidman},
  {Ellis}, {Irwin}, {McMahon}, {Ruiz-Lapuente}, {Walton}, {Schaefer}, {Boyle},
  {Filippenko}, {Matheson}, {Fruchter}, {Panagia}, {Newberg}, {Couch}, \&
  {Project}}]{Perlmutter}
{Perlmutter}, S., {Aldering}, G., {Goldhaber}, G., {et~al.} 1999, \apj, 517,
  565

\bibitem[{{Phillips}(1993)}]{Phillips}
{Phillips}, M.~M. 1993, \apjl, 413, L105

\bibitem[{{Piersanti} {et~al.}(2003){Piersanti}, {Gagliardi}, {Iben}, \&
  {Tornamb{\'e}}}]{2003ApJ...583..885P}
{Piersanti}, L., {Gagliardi}, S., {Iben}, Jr., I., \& {Tornamb{\'e}}, A. 2003,
  \apj, 583, 885

\bibitem[{{Potekhin} \& {Chabrier}(2000)}]{2000PhRvE..62.8554P}
{Potekhin}, A.~Y., \& {Chabrier}, G. 2000, \pre, 62, 8554

\bibitem[{{Pringle}(1974)}]{1974PhDT.......131P}
{Pringle}, J.~E. 1974, PhD thesis, , Univ.~Cambridge, (1974)

\bibitem[{{Raskin} {et~al.}(2012){Raskin}, {Scannapieco}, {Fryer},
  {Rockefeller}, \& {Timmes}}]{2012ApJ...746...62R}
{Raskin}, C., {Scannapieco}, E., {Fryer}, C., {Rockefeller}, G., \& {Timmes},
  F.~X. 2012, \apj, 746, 62

\bibitem[{{Riess} {et~al.}(1998){Riess}, {Filippenko}, {Challis},
  {Clocchiatti}, {Diercks}, {Garnavich}, {Gilliland}, {Hogan}, {Jha},
  {Kirshner}, {Leibundgut}, {Phillips}, {Reiss}, {Schmidt}, {Schommer},
  {Smith}, {Spyromilio}, {Stubbs}, {Suntzeff}, \& {Tonry}}]{Riess}
{Riess}, A.~G., {Filippenko}, A.~V., {Challis}, P., {et~al.} 1998, \aj, 116,
  1009

\bibitem[{{Rotondo} {et~al.}(2011){Rotondo}, {Rueda}, {Ruffini}, \&
  {Xue}}]{2011PhRvD..84h4007R}
{Rotondo}, M., {Rueda}, J.~A., {Ruffini}, R., \& {Xue}, S.-S. 2011, \prd, 84,
  084007

\bibitem[{{Saio} \& {Nomoto}(1985)}]{1985A&A...150L..21S}
{Saio}, H., \& {Nomoto}, K. 1985, \aap, 150, L21

\bibitem[{{Salpeter}(1961)}]{1961ApJ...134..669S}
{Salpeter}, E.~E. 1961, \apj, 134, 669

\bibitem[{{Sato} {et~al.}(2015){Sato}, {Nakasato}, {Tanikawa}, {Nomoto},
  {Maeda}, \& {Hachisu}}]{Sato15}
{Sato}, Y., {Nakasato}, N., {Tanikawa}, A., {et~al.} 2015, \apj, 807, 105

\bibitem[{{Schwab} {et~al.}(2012){Schwab}, {Shen}, {Quataert}, {Dan}, \&
  {Rosswog}}]{2012MNRAS.427..190S}
{Schwab}, J., {Shen}, K.~J., {Quataert}, E., {Dan}, M., \& {Rosswog}, S. 2012,
  \mnras, 427, 190

\bibitem[{{Shakura} \& {Sunyaev}(1973)}]{SS73}
{Shakura}, N.~I., \& {Sunyaev}, R.~A. 1973, \aap, 24, 337

\bibitem[{{Shen} {et~al.}(2012){Shen}, {Bildsten}, {Kasen}, \&
  {Quataert}}]{2012ApJ...748...35S}
{Shen}, K.~J., {Bildsten}, L., {Kasen}, D., \& {Quataert}, E. 2012, \apj, 748,
  35

\bibitem[{{Shibata} {et~al.}(2000){Shibata}, {Baumgarte}, \&
  {Shapiro}}]{2000PhRvD..61d4012S}
{Shibata}, M., {Baumgarte}, T.~W., \& {Shapiro}, S.~L. 2000, \prd, 61, 044012

\bibitem[{{Spitkovsky}(2006)}]{2006ApJ...648L..51S}
{Spitkovsky}, A. 2006, \apjl, 648, L51

\bibitem[{{Stergioulas}(2003)}]{2003LRR.....6....3S}
{Stergioulas}, N. 2003, Living Reviews in Relativity, 6, 3

\bibitem[{{Totani} {et~al.}(2008){Totani}, {Morokuma}, {Oda}, {Doi}, \&
  {Yasuda}}]{Totani}
{Totani}, T., {Morokuma}, T., {Oda}, T., {Doi}, M., \& {Yasuda}, N. 2008,
  \pasj, 60, 1327

\bibitem[{{Townsley} \& {Bildsten}(2004)}]{2004ApJ...600..390T}
{Townsley}, D.~M., \& {Bildsten}, L. 2004, \apj, 600, 390

\bibitem[{{van Kerkwijk} {et~al.}(2010){van Kerkwijk}, {Chang}, \&
  {Justham}}]{vk10}
{van Kerkwijk}, M.~H., {Chang}, P., \& {Justham}, S. 2010, \apjl, 722, L157

\bibitem[{{Wang}(1987)}]{1987A&A...183..257W}
{Wang}, Y.-M. 1987, \aap, 183, 257

\bibitem[{{Wang}(1995)}]{1995ApJ...449L.153W}
---. 1995, \apjl, 449, L153

\bibitem[{{Wang}(1997)}]{1997ApJ...475L.135W}
---. 1997, \apjl, 475, L135

\bibitem[{{Webbink}(1984)}]{Webbink1984}
{Webbink}, R.~F. 1984, \apj, 277, 355

\bibitem[{{Weinberg} {et~al.}(2013){Weinberg}, {Mortonson}, {Eisenstein},
  {Hirata}, {Riess}, \& {Rozo}}]{Weinberg}
{Weinberg}, D.~H., {Mortonson}, M.~J., {Eisenstein}, D.~J., {et~al.} 2013,
  \physrep, 530, 87

\bibitem[{{Yoon} {et~al.}(2007){Yoon}, {Podsiadlowski}, \&
  {Rosswog}}]{2007MNRAS.380..933Y}
{Yoon}, S.-C., {Podsiadlowski}, P., \& {Rosswog}, S. 2007, \mnras, 380, 933

\bibitem[{{Zhu} {et~al.}(2013){Zhu}, {Chang}, {van Kerkwijk}, \&
  {Wadsley}}]{2013ApJ...767..164Z}
{Zhu}, C., {Chang}, P., {van Kerkwijk}, M.~H., \& {Wadsley}, J. 2013, \apj,
  767, 164

\bibitem[{{Zhu} {et~al.}(2015){Zhu}, {Pakmor}, {van Kerkwijk}, \&
  {Chang}}]{2015ApJ...806L...1Z}
{Zhu}, C., {Pakmor}, R., {van Kerkwijk}, M.~H., \& {Chang}, P. 2015, \apjl,
  806, L1

\end{thebibliography}

\end{document}